# Supermassive black hole feeding and feedback observed on sub-parsec scales

Takuma Izumi[1,2,3,*], Keiichi Wada[4,5,6], Masatoshi Imanishi[1,3], Kouichiro Nakanishi[1,3], Kotaro Kohno[7,8], Yuki Kudoh[1,4,9], Taiki Kawamuro[10], Shunsuke Baba[4], Naoki Matsumoto[9], Yutaka Fujita[2], Konrad R. W. Tristram[11]

[1]National Astronomical Observatory of Japan, Tokyo 181-8588, Japan
[2]Department of Physics, Graduate School of Science, Tokyo Metropolitan University, Tokyo 192-0397, Japan
[3]Department of Astronomical Science, The Graduate University for Advanced Studies, SOKENDAI, Tokyo 181-8588, Japan
[4]Graduate School of Science and Engineering, Kagoshima University, Kagoshima 890-0065, Japan
[5]Research Center for Space and Cosmic Evolution, Ehime University, Ehime 790-8577, Japan
[6]Faculty of Science, Hokkaido University, Hokkaido 060-0810, Japan
[7]Institute of Astronomy, Graduate School of Science, The University of Tokyo, Tokyo 181-0015, Japan
[8]Research Center for the Early Universe, Graduate School of Science, The University of Tokyo, Tokyo 113-0033, Japan
[9]Astronomical Institute, Tohoku University, Miyagi 980-8578, Japan
[10]Cluster for Pioneering Research, RIKEN, Saitama 351-0198, Japan
[11]European Southern Observatory, Vitacura, Santiago 19001, Chile

*Corresponding author. Email: takuma.izumi@nao.ac.jp

**Active galaxies contain a supermassive black hole at their center, which grows by accreting matter from the surrounding galaxy. The accretion process in the central ~10 parsecs has not been directly resolved in previous observations, due to the small apparent angular sizes involved. We observed the active nucleus of the Circinus Galaxy using sub-millimeter interferometry. A dense inflow of molecular gas is evident on sub-parsec scales. We calculate that less than 3% of this inflow is accreted by the black hole, with the rest being ejected by multiphase outflows, providing feedback to the host galaxy. The observations also reveal a dense gas disk surrounding the inflow; the disc is gravitationally unstable which drives the accretion into the central ~1 parsec.**

Supermassive black holes (SMBHs), those with masses $M_{BH} > 10^6~M_\odot$ (where $M_\odot$ is the mass of the Sun), are ubiquitous in the centers of nearby galaxies (*1, 2*). SMBHs grow by accreting mass from the surrounding galaxy (*3*), which releases energy that can be observed as an active galactic nucleus (AGN) (*4*). Major mergers of gas-rich galaxies trigger intense SMBH accretion and bursts of star formation (*5, 6*). More gradual (secular) processes, such as perturbations caused by non-axisymmetric gravitational potentials and minor mergers (*7, 8*), are predicted to be sufficient to sustain lower luminosity AGNs, such as those classified as Seyfert galaxies.

Previous observations have shown that material can be efficiently supplied from the interstellar medium of an entire galaxy to its central region (the central ~100 parsecs [pc], about 1% of the size of the galaxy) (*9*). Resolved observations of this central region using molecular and atomic lines (*10-12*) have shown that the accumulated gas often forms a circumnuclear disk (CND) due to its retained angular momentum. Sub-millimeter observations have been used to study the parsec-scale molecular gas in nearby galaxies (*13, 14*). However, little is known about the supply of material to the innermost tens of pc (particularly within a radius from the center $r$ < 10 pc) in AGNs, where SMBHs usually dominate the gravitational potential (*2*). An exception is the nearby active galaxy NGC 1068, in which the subpc-scale $H_2O$ maser disk and the ~10 pc scale molecular torus have been found to counter-rotate, which was proposed to enhance mass accretion onto its SMBH (*15, 16*). However, similar dynamics have not been observed in other AGNs, so might not be common (*11*). Previous observations of cold gas have not provided quantitative estimates of the mass accretion rate at $r$ < 10 pc.

Absorption line measurements provide complementary information on the supply of gas to SMBHs, because they are sensitive to gas clouds along our line-of-sight toward the bright AGN. These can directly determine inflow and outflow velocities, from the red- and blue-shift of each line relative to the systemic velocity. Absorption line observations have shown that some AGNs have inflows of gas (*17, 18*). However, while some absorbers have been interpreted as located near to the central AGN ($r$ < 100 pc), this interpretation has been challenged (*19*). Those studies have been limited by the spatial resolution (~100 to 1000 pc) of the observations.

AGN-driven feedback impacts the star formation in the host galaxy (*20, 21*). The supply of material to the SMBH is known to trigger this feedback, though it has not been possible to quantitatively connect the gas supply to the subsequent feedback, especially the multiphase structure and dynamics of the gas at the parsec-scale around AGN (*9, 20*).

**Observations of the Circinus Galaxy**

To investigate parsec-scale feeding and feedback, we consider the Circinus Galaxy (also known as ESO 97-13), a nearby AGN (distance 4.2 ± 0.7 Mpc) (*22*). The active nucleus of this galaxy has a moderate intrinsic X-ray luminosity (at energies 2 to 10 kilo-electron-volt) of (2 to 5) × $10^{42}$ erg s$^{-1}$, which is heavily obscured by gas and dust (*23*). Around the SMBH, there is a sub-parsec scale gas disk, probed by $H_2O$ maser emission, that has been used to measure the dynamical mass of the SMBH: (1.7 ± 0.3) × $10^6$ M$_\odot$ (*24*). Other observations have shown a one-sided outflowing cone of ionized gas that extends for up to a kiloparsec (*25-27*). Previous submillimeter observations of carbon monoxide (CO) and atomic carbon ([C I]) showed a gas-rich CND ($r$ ~ 30 pc) with a denser gas concentration toward the inner 10 pc (*11, 28*). High-resolution ionized gas observations have been performed in the rest-frame optical and near-infrared wavelengths (*25-27*), but they are affected by dust extinction in the nucleus region.

We performed observations of the Circinus Galaxy with the Atacama Large Millimeter/submillimeter Array (ALMA) (*29*), at a spatial resolution of 0.5 to 2.6 pc. We supplement our observations with additional archival ALMA data. The combined dataset targeted emission lines of hydrogen cyanide HCN ($J$ = 3→2, where $J$ is the rotational quantum number) at rest frequency $\nu_{rest}$ = 265.8864 GHz, a recombination line of atomic hydrogen (H36α) at $\nu_{rest}$ = 135.2860 GHz, carbon monoxide CO ($J$ = 3→2) at $\nu_{rest}$ = 345.7960 GHz, and atomic carbon [C I] $^3P_1$–$^3P_0$ at $\nu_{rest}$ = 492.1607 GHz. For each line, we calculate the critical density $n_{cr}$, at which the molecule can be excited to the upper energy state of the line by collisions with $H_2$ molecules, in the optically thin limit as $n_{cr} = A_{ul}/C_{ul}$, where $A_{ul}$ and $C_{ul}$ are the Einstein A- and C-coefficients of a transition between upper state $u$ and lower state $l$. HCN ($J = 3\rightarrow 2$) traces very dense molecular gas ($n_{cr} = 5.8 \times 10^6$ cm$^{-3}$ at 50 K), so we expect it to probe parsec-scale dense flows more effectively than low-$J$ CO lines (which have $n_{cr} \sim 10^3$ to $10^4$ cm$^{-3}$). We expect the H36α line to trace feedback flows of ionized gas. The CO ($J = 3\rightarrow 2$) and [C I] $^3P_1$–$^3P_0$ lines trace medium-density molecular gas ($n_{cr} = 1.1 \times 10^4$ cm$^{-3}$ at 50 K) and diffuse atomic gas ($n_{cr} = 3.7 \times 10^2$ cm$^{-3}$), respectively. These submillimeter lines allow us to study the parsec-scale gas in multiple phases and densities, without being obscured by dust extinction.

Figure 1 shows the [C I] $^3P_1$–$^3P_0$ and CO ($J = 3\rightarrow 2$) observations; they trace the global spiral arms of the host galaxy, which converge to form the CND. Compared to CO ($J = 3\rightarrow 2$), [C I] $^3P_1$–$^3P_0$ is more concentrated in the central region (Fig. 1A), perhaps due to dissociation of CO into C by the X-ray radiation from the AGN (a central X-ray dominated region) (*29, 30, 31*). The map shows a parsec-scale compact concentration of HCN ($J = 3\rightarrow 2$) in the inner region (Fig. 1C; the central hole is due to continuum absorption). We expect such dense molecular gas to survive the X-ray irradiation because it is predominantly in the shielded mid-plane of the disk (the proposed geometrical structure is shown in Fig. 2E). Lower-density gas (traced by low-$J$ CO) is located above the mid-plane so is more easily dissociated. The parsec-scale distribution of H36α traces the ionized gas without severe dust extinction (Fig. 1D). A one-sided horn-like structure extends almost perpendicular to the HCN disk, toward the north-west (the middle position angle (PA) of the emission extended to the north-west is 313°, measured anti-clockwise from north). This direction, as well as the half-opening angle (~33° to 39°) of the H36α cone, are consistent with the larger kiloparsec-scale ionization cone seen in previous observations (*25, 26*).

**The parsec-scale inflow rate**

To estimate the mass inflow rate onto the SMBH, we first need to constrain the geometry of the disk. Figure 2A shows that the velocity pattern of the CND is dominated by rotation. We used this map to measure the position angles of the CND-scale kinematic major and minor axes, finding 210° and 300° respectively. The parsec-scale velocity pattern in the HCN ($J = 3\rightarrow 2$) map (Fig. 2B) indicates rotation following Keplerian orbits. This motion is evident within 0.1 arcsec (equivalent to 2 pc) from the SMBH in position-velocity diagrams (PVDs) of both HCN ($J = 3\rightarrow 2$) and [C I] $^3P_1$–$^3P_0$, measured along the CND-scale major axis (Figure 2C). This Keplerian motion is more prominent in the blue approaching side of the HCN disk than the red receding side, implying asymmetric gas mass distribution at the parsec-scale.

We modelled the three-dimensional structure of these multi-phase gases using a tilted-ring scheme (*32*) that decomposes the rotation velocity ($V_{rot}$) and random dispersion ($\sigma_{disp}$) (*29*). Figure 2D shows the decomposed rotation curves of CO ($J = 3\rightarrow 2$), [C I] $^3P_1$–$^3P_0$, and HCN ($J = 3\rightarrow 2$) in the central $r < 15$ pc. The rotation curves of CO ($J = 3\rightarrow 2$) and [C I] $^3P_1$–$^3P_0$ are consistent with each other, and smoothly connect to those previously measured over larger scales of $r = 10$ to 150 pc (*11*). The [C I] $^3P_1$–$^3P_0$ and HCN ($J = 3\rightarrow 2$) rotation curves indicate a Keplerian rotation pattern following a power law (velocity $\propto r^n$) with index $n = 0.46 \pm 0.06$. We estimate the SMBH mass from these curves, finding $(2.0 \pm 0.1) \times 10^6$ M$_\odot$, which is consistent with previous $H_2O$ maser observations (*24*).

Figure 2E shows radial profiles of the ratio $\sigma_{disp}/V_{rot}$ for each line, which are proxies for the aspect ratio (scale height to radius ratio) of the disk, assuming hydrostatic equilibrium. In the innermost region, within a few parsecs of the SMBH, the diffuse atomic gas (traced by [C I]) forms a geometrically thick structure, whereas the dense molecular gas (traced by HCN) is confined in a thin disk. This confirms our expectation that the dense molecular gas is predominantly in the mid-plane of the disk, where inflows toward the SMBH are expected (*33*).

We identify inflow motion in the HCN ($J = 3 \rightarrow 2$) spectrum measured at the position of the AGN (Fig. 3A), which displays a deep continuum absorption feature close to the systemic velocity ($V_{sys}$), as well as emission at velocities < 390 km s$^{-1}$. This absorption feature is skewed toward the redder side of $V_{sys}$, consistent with an inverse P-Cygni profile (a redder absorption and bluer emission pattern) which is characteristic of inflows. Because the dense gas forms a thin disk, and this disk is nearly edge-on (at least at the scales of the H$_2$O maser disk), we fitted this profile with a two-layer infall model (*29*, *34*), which indicates an inflow velocity of 7.4 ± 1.0 km s$^{-1}$. Assuming the volume-averaged gas density $n_{H2}$ equals the range of HCN ($J = 3 \rightarrow 2$) critical densities ($n_{cr}$ = (5.8 to 10) × 10$^6$ cm$^{-3}$ over a temperature range of 20 to 50 K), we estimate the mass inflow rate as ~0.20 to 0.34 M$_\odot$ yr$^{-1}$ at $r$ = 0.27 pc (half the spatial resolution). This inflow rate is sufficient to sustain the luminosity of the AGN [the SMBH accretion rate is ~0.006 M$_\odot$ yr$^{-1}$ (*29*)].

An alternative interpretation of the HCN ($J = 3 \rightarrow 2$) absorption is that it is caused by a dense cloud on an elliptical orbit around the SMBH. Such elliptical orbits have been frequently observed on larger galactic scales, but become less stable in the inner part of a galaxy because of the many overlapping gas cloud orbits. If the system is globally stable, frequent collisions and tidal interactions between gas clouds would disrupt the clouds. As a result, we expect a smooth circumnuclear disk with density fluctuations, with nearly circular rotation and self-regulated turbulence contributing to the viscosity of the gas (*35*). In the central parsec-scale region of the Circinus Galaxy, the expected mean free path of the clouds is much smaller than the disk size, so any clouds on elliptical orbits would collide with another cloud or be tidally-disrupted many times within even a single orbit (*29*). The HCN ($J = 3 \rightarrow 2$) rotation curve is consistent with Keplerian rotation, implying that most orbits in this region are circular. We see no evidence of a nuclear bar, which could introduce elliptical motions. Hence, we consider the elliptical orbit scenario to be unlikely, so conclude that the HCN ($J = 3 \rightarrow 2$) profile indicates inflow motion.

**Connection to larger scales**

The horn-like H36α distribution and its broad line width [the Gaussian full-width at half-maximum (FWHM) is 393 ± 44 km s$^{-1}$; Figure 3B] indicate a parsec-scale origin of the ionized outflow. The line profile of H36α is almost symmetric, indicating detection of both the receding and approaching sides of the ionization cone. In contrast, previous near-infrared coronal line profiles showed only blue tails, with velocities consistent with our H36α data (*27*); presumably because the receding side was obscured by dust in the earlier observations. The submillimeter H36α recombination line is much less affected by dust extinction, so probes both sides of the cone. Using the previously constrained electron temperature of the circumnuclear region (*29*) and assuming local-thermodynamic-equilibrium (LTE), we estimated the volume-averaged electron density of this region as ~7400 cm$^{-3}$ (*29*). We measured the outflow velocity from the line profile (~200 km s$^{-1}$), then used it to estimate the ionized outflow rate as ~0.04 M$_\odot$ yr$^{-1}$. The kinetic power of this outflow (~5 × 10$^{38}$ erg s$^{-1}$) corresponds to 0.001% of the AGN bolometric luminosity, considerably smaller than predicted by some AGN-feedback models, e.g. ~5% in the energy-conserving feedback model (*20*).

Our measured inflow and outflow rates imply that only small fractions of the inflowing mass are consumed by black hole growth (< 3%) or expelled in the ionized gas outflow (< 20%). Hence, the majority (> 80%) of the inflowing mass must be converted to outflows in other gas phases. The two [C I] $^3P_1$–$^3P_0$ peaks in the minor-axis PVD (offset ~ 0.1 arcsec, equivalent to 2 pc), where CO ($J = 3\rightarrow 2$) becomes fainter, are offset from the $V_{sys}$ (446 km s$^{-1}$) by ~15 km s$^{-1}$ (Fig. 3C). Considering the orientation of the Circinus Galaxy [southeast is the near side (*11*)], these offset components indicate outflows of atomic gas. The faint CO emission in that location might be due to weak molecular outflows. The maximum [C I] $^3P_1$–$^3P_0$ outflow velocity (~40 km s$^{-1}$) is slower than the escape velocity from the SMBH gravitational potential at this scale (*29*). Therefore, we expect these outflows visible in [C I] $^3P_1$–$^3P_0$ to stall and become backflows to the CND, causing additional turbulence that could support the geometrical thickness of the atomic gas disk (*33*). Figure 3D summarizes the multiphase gas flows we identify.

**Physical process driving the accretion**

To determine the physical process responsible for the accretion seen in the HCN ($J = 3\rightarrow 2$) profile, we first considered turbulent viscosity in the dense gas disk. The turbulent accretion timescale is $t_{acc} = r^2/v_{vis}$, where $v_{vis}$ is the viscous parameter, defined as $v_{vis} = \alpha\,\sigma_{disp}\,h$, where $h$ is the disk scale height and $\alpha$ is an efficiency parameter in the range $0 < \alpha < 1$ (*36*). This gives an accretion timescale of $1.8 \times 10^5 \times \alpha^{-1}$ yr at $r = 0.27$ pc (taking $\sigma_{disp} \sim 15$ km s$^{-1}$ and $h \sim 0.1 \times r$, estimated from Fig. 2E). The enclosed dense gas mass within this radius is $\pi r^2 \times 2h \times 2n_{H2} \times m_p \sim$ 3500 to 6100 M$_\odot$, again assuming that $n_{H2}$ spans the range of HCN ($J = 3\rightarrow 2$) critical densities given above. The mass inflow rate expected under this viscous accretion is then $\sim \alpha \times$ (0.02–0.03) M$_\odot$ yr$^{-1}$. Even for the highest possible value of $\alpha=1$, this inflow rate is an order of magnitude smaller than we deduced from the infall model. We therefore find viscous accretion to be an unlikely mechanism at this scale of several parsecs.

Alternatively, the torques induced by non-axisymmetric gravitational instabilities [caused by spiral arms or clumps (*37*)] or gas clump-clump collisions (*7*) are possible mechanisms for angular momentum transfer. We estimated the radial profile of the Toomre $Q$-parameter, which expresses the balance between the self-gravity of the molecular gas and turbulent pressure (*38*). $Q \equiv \kappa c_s/\pi G\Sigma$, where $\kappa$, $c_s$, $G$, and $\Sigma$ are the epicyclic frequency of the orbit, the speed of sound, the gravitational constant, and the surface density of the gas mass, respectively. A geometrically thin disk becomes gravitationally unstable if $Q < 1$. We assumed $\Sigma = 2n_{H2} \times 2h \times m_p$ and the same range of $n_{H2}$ as above. Substituting $\sigma_{disp}$ into the expression for $c_s$ and taking $h = (\sigma_{disp}/V_{rot}) \times r$, we obtained the profiles shown in Figure 4. The derived values of $Q$ are less than the critical value at $r > 1$ pc, indicating that the disk is gravitationally unstable, so we expect clump-clump collisions or instability-driven turbulence (*39*) to occur. For the regions slightly above $Q = 1$, we expect non-axisymmetric structures (such as a nuclear spiral arm) to be induced, which would also cause a gravitational torque. Some non-axisymmetric structures are visible in the HCN disk (Fig. 1C).

In contrast, the $Q > 1$ at $r < 1$ pc (Figure 4). A dense gas mass of $> 3 \times 10^6$ M$_\odot$ would be required to make the disk gravitationally unstable at $r < 0.4$ pc if the disk scale height is $h \sim 0.1 \times r$, as we adopted above. We rule out such a high mass of dense gas based on our HCN ($J = 3\rightarrow 2$) rotation curve, which indicates that the enclosed mass is $\sim 2 \times 10^6$ M$_\odot$ within $r = 1$ pc. Kinematic viscosity due to magnetohydrodynamic turbulence would not be sufficient; numerical models of that process predict $\alpha \sim 10^{-4}$ *(40)*. To drive accretion on sub-parsec scales would require a further denser ($>10^8$ cm$^{-3}$) but thin disk, which might correspond to the disk seen in the H$_2$O maser observations (*24*). Our HCN ($J = 3 \rightarrow 2$) observations would not be sensitive to such a disk, but it is theoretically capable of driving the gravitational instability; we therefore prefer this explanation.

**Acknowledgments:** We thank the anonymous reviewers for their thorough reading and constructive comments. ALMA is a partnership of ESO (representing its member states), NSF (USA), and NINS (Japan), together with NRC (Canada), MOST and ASIAA (Taiwan), and KASI (Republic of Korea), in cooperation with the Republic of Chile. The Joint ALMA Observatory is operated by ESO, AUI/NRAO, and NAOJ.

**Funding:** Funded by the Japan Society for the Promotion of Science (JSPS) through KAKENHI Grant Numbers: JP20K14531 (T.I.); JP21H04496 (K.W.); JP17H06130 (K.K.); JP21K03632 (M.I.); JP19K03937 (K.N.); JP20K14529 (T.K.); and JP20H00181, JP22H00158 & JP22H01268 (all Y.F.). Also funded by NAOJ ALMA Scientific Research Grant Codes 2020-14A (Y.K., K.W.) and 2022-21A (T.I., Y.F.), and by ALMA Japan Research Grant for the NAOJ ALMA Project code NAOJ-ALMA-271 (T.I.).

**Author contributions:** T.I. led the project, analyzed the data, produced figures, and led the manuscript writing. K.W., Y.K., and S.B. contributed to the theoretical interpretation and produced figures. All authors contributed to the scientific discussion and writing.

**Competing interests:** The authors declare that they have no competing interests.

**Data and materials availability:** The ALMA data are available from the ALMA archive at https://almascience.nao.ac.jp/aq/ . This paper makes use of the following ALMA data: ADS/JAO.ALMA #2019.1.00014.S, #2017.1.00575.S, and #2018.1.00581.S. Our reduced ALMA data cubes, dynamical model input, and dynamical model results are archived at the UTokyo Repository (*41*).

**Supplementary Materials**

Materials and Methods

Figs. S1 to S9

Tables S1 to S2

References (*42–66*)

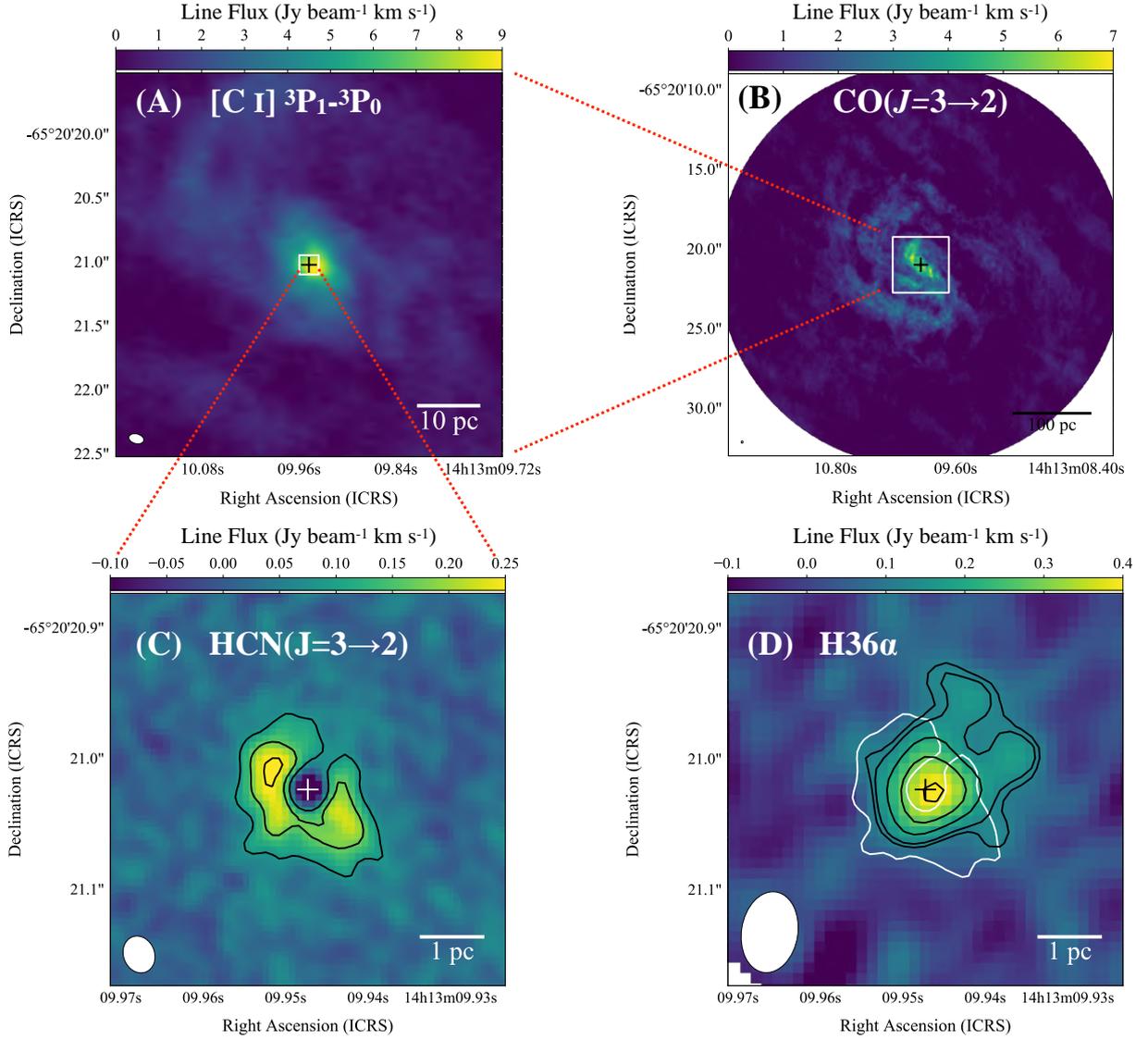

**Figure 1. Observed spatial distributions of multiple gas phases. (A)** Atomic gas traced by [C I] $^3P_1$–$^3P_0$ (uncertainty $1\sigma = 0.088$ Jy beam$^{-1}$ km s$^{-1}$) in the central 7 arcsec (~140 pc) of the Circinus Galaxy. The white box indicates the region shown in panels C and D. **(B)** Medium density gas traced by CO ($J = 3\rightarrow2$) ($1\sigma = 0.055$ Jy beam$^{-1}$ km s$^{-1}$) on larger scales. The white square indicates the region shown in panel A. **(C)** Dense molecular gas traced by HCN ($J = 3\rightarrow2$) ($1\sigma = 0.017$ Jy beam$^{-1}$ km s$^{-1}$; contours are drawn at $5\sigma$, $10\sigma$, and $15\sigma$). **(D)** Ionized gas traced by H36α (black contours, drawn at $3\sigma$, $3.5\sigma$, $5\sigma$, $7\sigma$, and $10\sigma$, where $1\sigma = 0.040$ Jy beam$^{-1}$ km s$^{-1}$). The white contour is the $5\sigma$ contour of HCN ($J = 3\rightarrow2$) from panel C. In all panels, the plus sign indicates the AGN position, and the ellipses at the bottom left indicate the synthesized beam sizes. The coordinates are expressed in the International Celestial Reference System (ICRS).

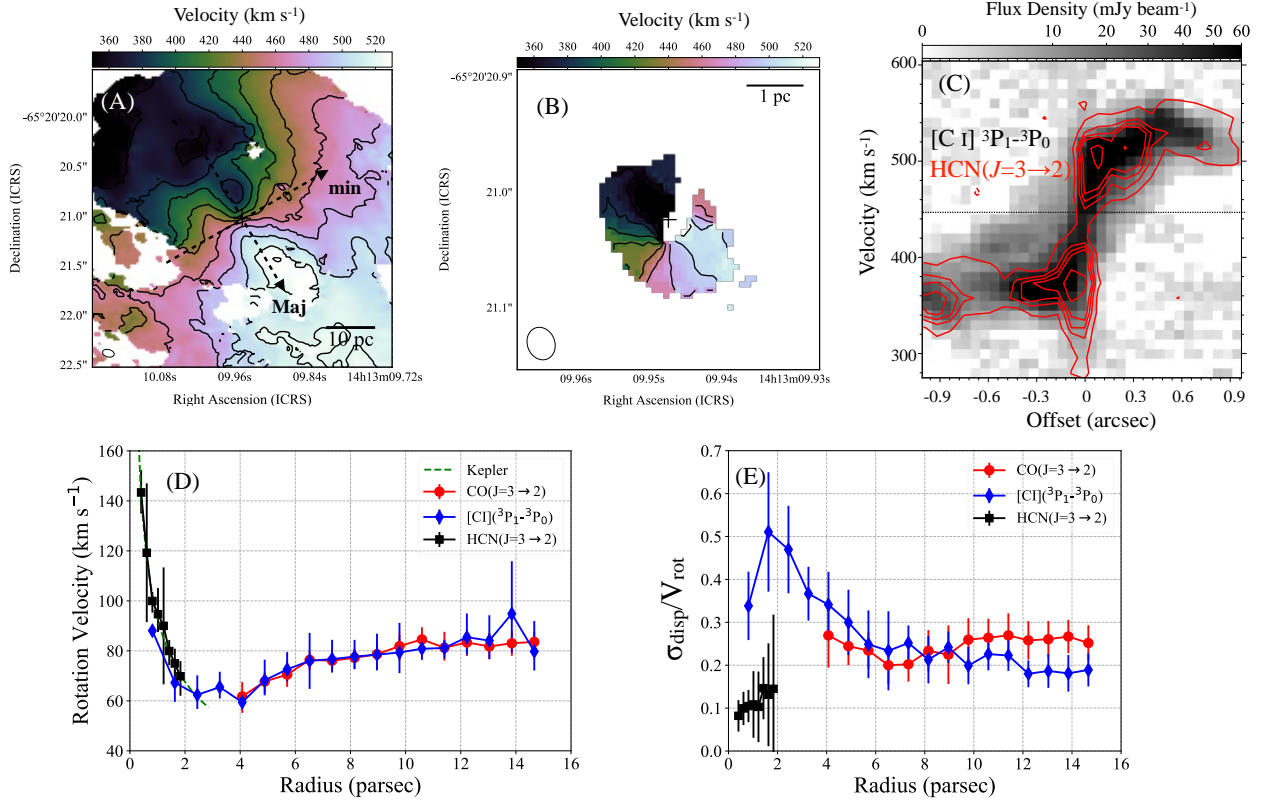

**Figure 2. Kinematic properties of the observed gas.** **(A)** Line-of-sight velocity (color bar) measured from the [C I] $^3P_1$–$^3P_0$ data, with dashed arrows indicating the kinematic major and minor axes as labelled. Fig. S3A shows an equivalent plot for CO ($J = 3 \rightarrow 2$). **(B)** Parsec-scale velocity field measured from the HCN ($J = 3 \rightarrow 2$) data. The absorption component close to the AGN was masked out. In panels A and B, black contours start at 360 km s$^{-1}$ and are separated by steps of 20 km s$^{-1}$. Black ellipses at the lower left indicate the beam sizes. **(C)** Major-axis PVDs of [C I] $^3P_1$–$^3P_0$ (grayscale, 1$\sigma$ = 1.26 mJy beam$^{-1}$) and beam-matched HCN ($J = 3 \rightarrow 2$) (red contours at –3$\sigma$, 3$\sigma$, 5$\sigma$, 6$\sigma$, 7$\sigma$, and 10$\sigma$, where 1$\sigma$ = 0.68 mJy beam$^{-1}$). The horizontal dashed line indicates the systemic velocity of 446 km s$^{-1}$. **(D)** Rotation curves of each of the gas tracers, as indicated in the legend. The green dashed line is a model of Keplerian motion due to a central black hole of mass $2.0 \times 10^6$ M$_\odot$, which was fitted to the data at $r < 2$ pc. **(E)** Radial profile of the $\sigma_{disp}/V_{rot}$ ratio, which is a proxy for the disk aspect ratio, for the same gas tracers. Near the AGN, the HCN ($J = 3 \rightarrow 2$) disk is much thinner than the [C I] $^3P_1$–$^3P_0$ disk. In panels (D) and (E), error bars indicate 1$\sigma$ uncertainty.

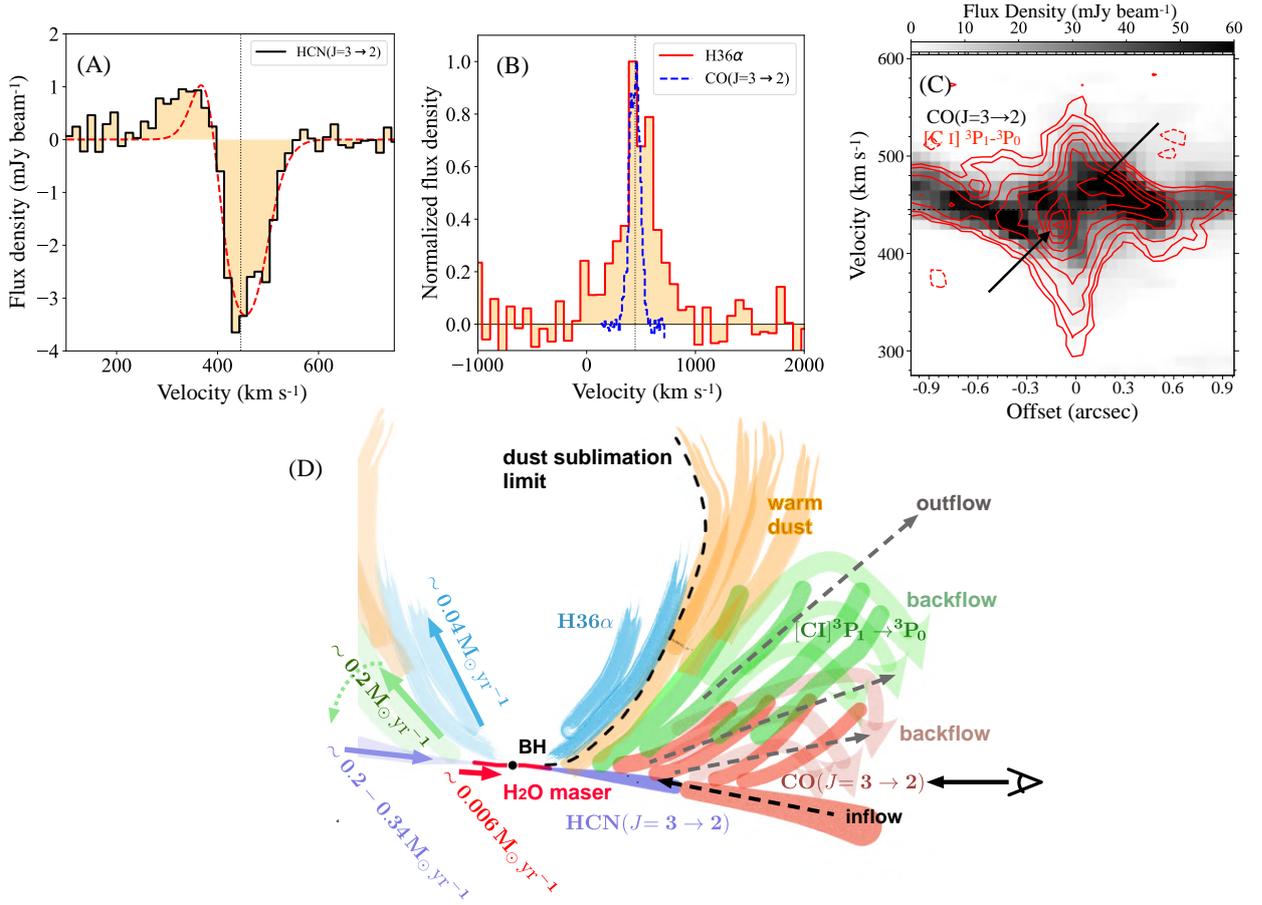

**Figure 3. Spectra at the AGN position and our inferred geometry of the multiphase gas flows. (A)** HCN ($J = 3 \rightarrow 2$) spectrum (black histogram with orange shading) measured at the AGN position, overlain with an infall model (dashed red line) (*29*). **(B)** H36α (red histogram with orange chading) and CO ($J = 3 \rightarrow 2$) (blue dashed line) spectra at the same position, normalized by their peak fluxes. The Gaussian FWHM is 393 ± 44 km s$^{-1}$ for H36α and 115 ± 1 km s$^{-1}$ for CO ($J = 3 \rightarrow 2$). In panels A and B, vertical dotted lines indicate $V_{sys.}$. **(C)** Minor-axis PVDs of CO ($J = 3 \rightarrow 2$) (grayscale, 1σ = 0.78 mJy beam$^{-1}$) and [C I] $^3P_1$–$^3P_0$ (red contours drawn at −3σ, 3σ, 5σ, 10σ, 20σ,…, 90σ, where 1σ = 1.26 mJy beam$^{-1}$). The black arrows indicate outflow features discussed in the text. **(D)** Schematic diagram showing a cross-section of the geometry we infer for the parsec-scale multiphase gas flows. The large scale spiral arms converge to form the CND we observe in CO ($J = 3 \rightarrow 2$) (thick red bands). At the central parsec scale, there is a dense molecular gas disk (purple bands) traced by HCN ($J = 3 \rightarrow 2$), which shows inflow motion (black dashed arrow). This dense disk connects to the H$_2$O maser disk at the central sub-parsec scale (thin red line). Fast ionized outflows observed in H36α (blue bands) and slow atomic outflows observed in [C I] $^3P_1$–$^3P_0$ (green bands) occur above the disk. The atomic outflow is slower so will fall back to the disk (labelled 'backflow'). Previously reported mid-infrared dust elongation toward the polar direction (*33*) is interpreted as dust entrained by these winds (orange bands). The dense molecular disk is gravitationally unstable in the inner parsec (Fig. 4), which drives the accretion and consequent feedback. Estimated mass flow rates in each phase (colored arrows) are labelled on the left of the panel. The black arrow indicates our line-of-sight.

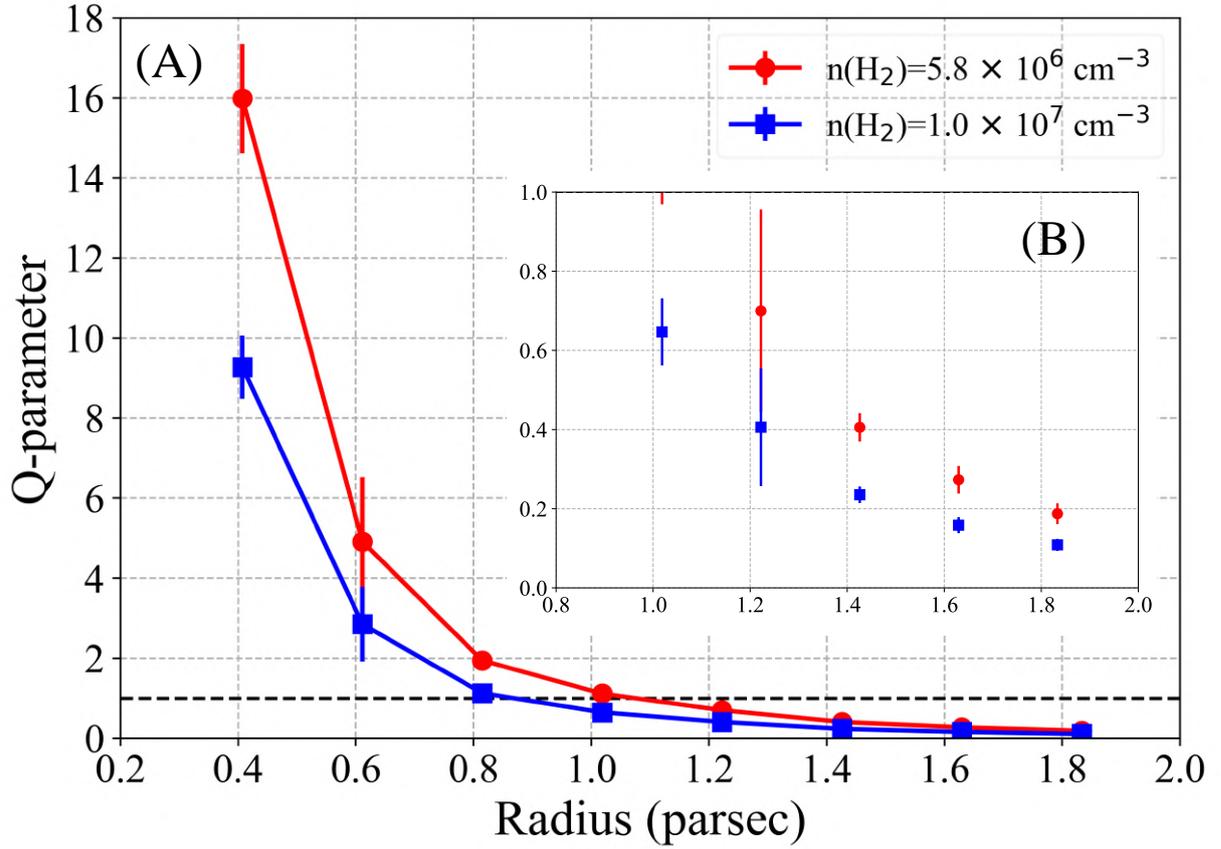

**Figure 4. Gravitational stability of the dense gas disk.** (**A**) Radial profiles of the Toomre-$Q$ parameter, plotted for two assumed values of the gas volume density (see legend). The horizontal dashed line indicates the critical value for stability, $Q=1$. Error bars indicate 1σ uncertainty. (**B**) Same as panel A, but zoomed on the region $r > 0.8$ pc and $Q<1$.

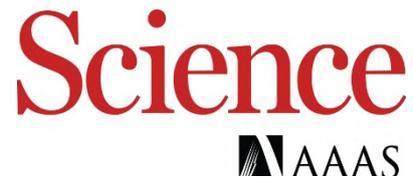

Supplementary Materials for

# Supermassive black hole feeding and feedback observed on sub-parsec scales

Takuma Izumi*, Keiichi Wada, Masatoshi Imanishi, Kouichiro Nakanishi, Kotaro Kohno, Yuki Kudoh, Taiki Kawamuro, Shunsuke Baba, Naoki Matsumoto, Yutaka Fujita, Konrad R. W. Tristram

*Correspondence to: takuma.izumi@nao.ac.jp

**This PDF file includes:**
    Materials and Methods
    Figs. S1 to S9
    Tables S1 to S2

**Materials and Methods**
**I. ALMA observations and data reduction**
The data used in this study contained three distinct observation programs that were conducted in various ALMA cycles (from 2017 to 2021). The resultant cube parameters are listed in Table S1. The absolute flux uncertainty is typically ~10%.

**HCN($J = 3 \rightarrow 2$) observations**
We analyzed archival observations of the Circinus Galaxy (project ID: 2018.1.00581. S, principal investigator (PI) K. Tristram) containing the HCN($J = 3 \rightarrow 2$) line at the rest frequency $v_{rest}$ = 265.8864 GHz. The observations were performed with ALMA Band 6 (211 to 275 GHz) during June and August 2018, using 41 to 46 antennas. Three array configurations were used with combined baseline lengths ranging from 41 m to 15238 m. The corresponding maximum recoverable scale (MRS) was 1.4 arcsec. We applied the Common Astronomy Software Applications (CASA) (*42*) task `tclean` to the pipeline-calibrated data (Briggs weighting, robust parameter = +0.5), which yielded a synthesized beam of 0.029 arcsec × 0.024 arcsec, equivalent to 0.6 pc × 0.5 pc (PA = 20.3°) at our adopted distance to the Circinus Galaxy. The 1σ channel sensitivity is 0.20 mJy beam$^{-1}$ at the velocity resolution $dV$ = 15 km s$^{-1}$. The pixel scale for the HCN($J = 3 \rightarrow 2$) maps is 0.005 arcsec. We imaged the underlying 1.1 mm continuum emission, which was used to model the absorption profile. This continuum emission was subtracted in the uv-plane before making the line cube. For this continuum map, the synthesized beam is 0.026 arcsec × 0.022 arcsec, equivalent to 0.5 pc × 0.4 pc with PA = 20.5°. Owing to the substantial brightness of the emission, we performed a two-round phase-only self-calibration to improve the dynamic range. Consequently, the resultant 1σ sensitivity of this continuum map is 14.7 μJy beam$^{-1}$, and the continuum peak position is (right ascension, declination) = (14$^h$13$^m$09$^s$.9473, −65°20′21″.024) in ICRS, which we adopt as the AGN position.

**H36α observations**
We analyzed another archival ALMA dataset (project ID: 2017.1.00575. S, PI J. Zhang) containing the H36α ($v_{rest}$ = 135.2860 GHz) line. Although this recombination line is adjacent to H$_2$CS (4$_{1,4}$–3$_{1,3}$) at $v_{rest}$ = 135.2984 GHz, we expect contamination to be negligible, because the innermost parsec-scale is influenced by X-ray radiation (destroying complex molecules). The observations were conducted in October 2017 with ALMA Band 4 (125 to 163 GHz) with 45 antennas, and two configurations were used with combined baseline lengths ranging from 41 to 16196 m and MRS of 1.4 arcsec. The H36a line spans two contiguous spectral windows—one spectral window (hereinafter, denoted as H36α-high data) covers the line-of-sight velocity of $V_{LOS}$ > 120 km s$^{-1}$ and the other (H36α-low data) covers $V_{LOS}$ < 200 km s$^{-1}$—both displaying consistent flux levels in the overlapping velocity range. The CASA `tclean` task for the pipeline-processed data (natural weighting to maximize the point source sensitivity) yielded a synthesized beam of 0.062 arcsec × 0.043 arcsec, equivalent to 1.3 pc × 0.9 pc (PA = -8.8°) for H36α-high, and 0.058 arcsec × 0.041 arcsec, equivalent to 1.2 pc × 0.8 pc (PA = –0.2°) for H36α-low. Given the faintness of the line, we degraded $dV$ to 75 km s$^{-1}$, giving a 1σ channel sensitivity of 0.18 mJy beam$^{-1}$. The pixel scale of the map is 0.008 arcsec. To extract the spectrum illustrated in Figures 3 and S1, we convolved the two cubes to a common circular resolution of 0.065 arcsec. The resulting 1σ sensitivity is 0.20 mJy beam$^{-1}$.

**CO($J = 3 \rightarrow 2$) and [C I] $^3P_1$–$^3P_0$ observations**
We observed the Circinus Galaxy between 2019 and 2021 with ALMA Bands 7 (275 to 373 GHz) and 8 (385 to 500 GHz) using 35 to 50 antennas (project ID = 2019.1.00014.S; PI

= T. Izumi). The target lines were CO($J$ = 3→2) at $v_{\rm rest}$ = 345.7960 GHz and [C I] $^3P_1$–$^3P_0$ at $v_{\rm rest}$ = 492.1607 GHz. The observations were conducted at a single pointing with fields of view of 17 arcsec (Band 7) and 12 arcsec (Band 8), which completely covered the CND (diameter ~ 2 arcsec). After combining the data acquired with various array configurations, the MRSs of these observations are greater than 1.5 arcsec (30 pc). The phase-tracking center was set to the previously observed 860 μm continuum peak position, i.e., (right ascension, declination) = (14$^{\rm h}$13$^{\rm m}$09$^{\rm s}$.948, –65°20 ′21″.05) in ICRS *(11)*. The combined baseline lengths ranged from 15 to 2517 m for CO($J$ = 3→2) and 15 to 3638 m for [C I] $^3P_1$–$^3P_0$, which produced a final synthesized beam size of 0.140 arcsec × 0.117 arcsec, equivalent to 2.9 pc × 2.4 pc (PA = –14.4°) for CO($J$ = 3→2) and 0.119 arcsec × 0.076 arcsec, equivalent to 2.4 pc × 1.5 pc for [C I] $^3P_1$–$^3P_0$, respectively. The line and underlying continuum emissions were reconstructed using the CASA `tclean` task with Briggs weighting (robust parameter = +0.2) down to 3σ level. The 1σ channel sensitivities were determined at channels free of line emission, and we obtained 1σ = 0.78 mJy beam$^{-1}$ and 1.26 mJy beam$^{-1}$ for CO($J$ = 3→2) and [C I] $^3P_1$–$^3P_0$ at $dV$ = 10 km s$^{-1}$, respectively. The pixel scales are set to 0.02 arcsec for these lines.

**Line profiles**
The line profiles of CO($J$ = 3→2), [C I] $^3P_1$–$^3P_0$, and H36α measured at the AGN position are shown in Figure S1. For the CO($J$ = 3→2) and [C I] $^3P_1$–$^3P_0$, we convolved the cubes to a common 0.14 arcsec resolution to take a line ratio. We fitted a single Gaussian function to the spectra to extract the basic parameters (Table S2). The resulting line center velocities of CO and [C I] were consistent with the previously measured value of 446 km s$^{-1}$ *(11)*. Thus, we set this 446 km s$^{-1}$ as the systemic velocity ($V_{\rm sys}$) to maintain consistency with prior studies. This $V_{\rm sys}$ was fixed when fitting the noisy H36α profile. The HCN($J$ = 3→2) profile, which contains an absorption feature, is plotted in Figure 3.

We fitted the CO($J$ = 3→2) profile with a single Gaussian function, finding that simple rotation is the dominant motion of this medium-density molecular gas. However, the [C I] $^3P_1$–$^3P_0$ profile is wider as well as skewed in comparison to the CO($J$ = 3→2) profile. Therefore, this atomic gas is more turbulent and traces multiple flows, such as outflows, in addition to rotation. Similar skewness in the [C I] $^3P_1$–$^3P_0$ profile has previously been found on a larger ~15 pc scale *(11)*, which was attributed to atomic outflows. The H36α profile is considerably wider than both CO($J$ = 3→2) and [C I] $^3P_1$–$^3P_0$ profiles, indicating that this line traces much faster gas motion (i.e., fast ionized outflow) as discussed in the main text. As there could be two distinct components in the vicinity of the AGN (i.e., compact core and extended outflow) traced by this H36α emission, we attempted to fit the line profile with a double-Gaussian function. However, we found this to be impractical due to the limited signal-to-noise ratio, so adopted single-Gaussian fits. The line profile extends to ±500 km s$^{-1}$ relative to $V_{\rm sys}$ toward both the blue and red sides. This blue-side extent is consistent with that measured from near-infrared coronal emission lines arising from the outflows *(27)*, indicating that H36α traces the innermost portion of the ionized outflows of the Circinus Galaxy. However, we also detected a red component. The coronal emission lines do not display this extent toward the red side, most likely because of dust extinction on the redder region (i.e., spatially far side) of the outflows.

Upon inspecting Figure 3 and S1, we adopted the velocity ranges of [200, 680], [200, 800], [200, 680], and [200, 680] km s$^{-1}$ to produce velocity-integrated intensity maps of HCN($J$ = 3 →2), H36α, CO($J$ = 3→2), and [C I] $^3P_1$–$^3P_0$, respectively (Figure 1).

## II. X-ray Dominated Region (XDR)

Compared to CO($J = 3\rightarrow2$), [C I] $^3P_1$–$^3P_0$ is more concentrated in the central region (Figure 1A). This trend is inconsistent with a typical increasing gas density profile toward the center of a galaxy, given the much higher $n_{cr}$ of CO($J = 3\rightarrow2$) than [C I] $^3P_1$–$^3P_0$. A higher gas temperature, as expected around the AGN, does not resolve this discrepancy, because the level energy for CO($J = 3\rightarrow2$) is higher than that for [C I] $^3P_1$–$^3P_0$. We found the [C I] $^3P_1$–$^3P_0$/CO($J = 3\rightarrow2$) brightness temperature line ratio is 1.2 ± 0.1 at the AGN position, using the line fluxes in Table S2. This is higher than the typical value detected in the star-forming regions of galaxies (~0.2) (*31, 43, 44*), which we interpret as due to AGN-triggered dissociation of molecules at the nucleus of the Circinus Galaxy. This [C I]-to-CO line ratio is 2.5 times greater than that observed at the central ~100 pc of NGC 7469 (luminous type-1 AGN) (*31*). Our parsec-resolution observations spatially resolve the regions of molecular dissociation, i.e., XDR (*30*) of the Circinus Galaxy, by reducing the contamination of the surrounding star-forming regions. A central concentration of [C I] $^3P_1$–$^3P_0$ relative to CO($J = 3\rightarrow2$) is also observed in the velocity channel maps (Figure S2-S4). Previous chemo-hydrodynamic 3D radiative transfer simulations (*45*) showed that CO($J = 3\rightarrow2$) line becomes optically thick near the disk plane (when viewed nearly edge-on). However, at high latitude above the disk, the line is optically thin. Hence the effect of optical depth (specifically self-absorption) does not affect our conclusions.

In XDRs, the fast electrons produced by the primary X-ray ionization cause further secondary ionization, efficient gas heating, and photodissociation. We quantify these effects using the effective ionization parameter (*29*),

$$\xi_{\text{eff}} = 1.26 \times 10^{-4}\, F_X\, n_5^{-1}\, N_{22}^{-\Phi}, \qquad \text{(S1)}$$

where $F_X$ denotes the incident 1–100 keV flux in units of erg s$^{-1}$ cm$^{-2}$, $n_5$ indicates the gas volume density in units of $10^5$ cm$^{-3}$, $N_{22}$ denotes the attenuating column density in units of $10^{22}$ cm$^{-2}$. The parameter $\Phi$ is related to the photon index ($\Gamma$) of the X-ray spectral energy distribution (SED) by $\Phi = (\Gamma + 2/3)/(8/3)$. According to XDR chemical calculations (*30*), $\xi_{\text{eff}}$ must be greater than $10^{-2.5}$ to increase the C abundance from that of CO in dense gas ($n_{H2}$ ~$10^5$ cm$^{-3}$). The 1–100 keV luminosity (~(0.5–1) × $10^{43}$ erg s$^{-1}$) and $\Gamma$ of the Circinus Galaxy have been estimated from wide-band X-ray SED models (*23*). Assuming typical values measured in the CNDs of nearby galaxies, $n_5 = 1$ and $N_{22} = 10$ (*31, 46, 47*), we find $r < 10$–15 pc satisfies $\xi_{\text{eff}} > 10^{-2.5}$. Although this is a rudimentary estimate, this spatial extent is consistent with the size of the [C I]-enhanced region of the Circinus Galaxy.

## III. Velocity maps

We integrated the same velocity range of [200, 680] km s$^{-1}$ to produce the line-of-sight velocity (moment-1) and velocity dispersion (moment-2) maps of HCN($J = 3\rightarrow2$), CO($J = 3\rightarrow2$) and [C I] $^3P_1$–$^3P_0$ after 5σ (HCN) and 10σ (CO + C I) clippings to avoid noise contamination (Figure 2 and S5). The global line-of-sight velocity patterns of CO and C I are consistent, both indicating galactic rotation. The velocity pattern of HCN($J = 3\rightarrow2$) instead traces the nuclear Keplerian motion.

The CO($J = 3\rightarrow2$) velocity dispersion map displays a highly inclined disk-like distribution along the CND, which we ascribe to the turbulent nature of the CND. However, we observed faint components close to $V_{sys}$ at a spatial offset > 0.3 arcsec in the major axis PVD, particularly in CO($J = 3\rightarrow2$) (Figure S5; at spatial offset < –0.3 arcsec, both CO ($J = 3\rightarrow2$) and [C I] $^3P_1$–$^3P_0$ have faint components close to $V_{sys}$). If the random motion contributes to the faint components close to $V_{sys}$, we should observe a broadening of the PVD also in the opposite direction from $V_{sys}$. However, we did not observe emission at $V_{LOS} \sim 600$ km s$^{-1}$ at spatial offset > 0.3 arcsec and ~250 km s$^{-1}$ at spatial offset < -0.3 arcsec. Therefore, random turbulent motion cannot explain these results. Instead, we interpret these components at $\sim V_{sys}$ as inflows of molecular CO, but not predominantly atomic C. This inflow would continue to the central parsec, with denser gas, which is probed by HCN($J = 3 \rightarrow 2$) absorption (as discussed in the main text). This inflow also contributes to the higher dispersion of CO($J = 3 \rightarrow 2$) than [C I] $^3P_1$–$^3P_0$. The spatial distribution of the high [C I] $^3P_1$–$^3P_0$ dispersion region is distinct from that of CO($J = 3 \rightarrow 2$), with a chimney-like extension toward the northwest. We interpret this extension as related to slow-velocity outflows (see below).

## IV. Three-dimensional dynamical modeling

Beam-deconvolved dynamical information was obtained by fitting concentric tilted rings to the data cubes. We used the $^{3D}$BAROLO code (*32*). The major parameters include the dynamic center $V_{sys}$, rotation velocity ($V_{rot}$), local velocity dispersion ($\sigma_{disp}$), radial velocity ($V_{rad}$), inclination angle (*i*), and position angle (*PA)*, all of which were allowed to vary in each ring (*11, 31, 48, 49*). For CO($J = 3 \rightarrow 2$) and [C I] $^3P_1$–$^3P_0$, we set the dynamical center to the AGN position, $V_{sys}$ as 446 km s$^{-1}$, and PA as 210° (global kinematic major axis; Figure 2A), to improve convergence. We set $V_{rad} = 0$ km s$^{-1}$, so any noncircular motion should be apparent in the residual map, produced by subtracting the model from the observed velocity field (Figures 2A, 2B, S5A). A combination of the Keplerian motion surrounding the (previously constrained) $1.7 \times 10^6$ M$_\odot$ SMBH (*24*) and the extrapolation of the rotation curve constructed at $r > 10$ pc (*11*) are used as our initial guesses for $V_{rot}$. The initial guesses for $\sigma_{disp}$ and $i$ were 15 km s$^{-1}$ and 75°, based on previous dynamical models for the Circinus galaxy (*11*). The models were constructed with $r = 0.04, 0.08, …, 0.72$ arcsec concentric rings to minimize the residual amplitude.

For the HCN($J = 3 \rightarrow 2$) model, we changed the PA to 225°, judged from the observed velocity field (Figure 2B), and fixed $i$ to 82° (the value obtained from our CO and [C I] models), given the faintness of this line. By inspecting the channel maps (Figure S4) and PVD (Figure 2C), we found that the fast Keplerian motion is visible in the blue, approaching side of the HCN disk up to −170 km s$^{-1}$, while it is not prominent in the red receding side (emission is visible only up to +100 km s$^{-1}$). However, we expect fast red components as seen in the H$_2$O maser disk (*24*). We thus interpret this asymmetry as due to asymmetric gas mass distribution inside the parsec-scale dense gas disk. Because the velocity extent seen in the blue side coincides with that seen in the H$_2$O maser disk, we only used this blue approaching side for the tilted ring model fitting. This HCN($J = 3 \rightarrow 2$) model was constructed from $r = 0.02$ arcsec ring to 0.09 arcsec, in intervals of 0.01 arcsec. The central $r < 0.02$ arcsec region was masked out, due to the absorption feature.

The model and residual maps are illustrated in Figure S6, and the resulting dynamic parameters are plotted in Figure S7. We find an upturn of $V_{rot}$ toward the center in these lines. The values of [C I] $^3P_1$–$^3P_0$ and HCN($J = 3 \rightarrow 2$) are consistent with those measured with H$_2$O maser observations (*24*), indicating that these are caused by the Keplerian rotation surrounding the SMBH. However, at the innermost $r < 4$ pc, $V_{rot}$ of CO($J = 3 \rightarrow 2$) is smaller than the other tracers, below the values required for Keplerian rotation. A possible explanation for this behavior is that the CO($J = 3 \rightarrow 2$) emission is optically thick at this innermost location, where the gas disk is nearly edge-on; this would cause emission to be self-absorbed by the foreground gas at $r \gtrsim 10$ pc. This is predicted by non-local thermodynamic equilibrium (non-LTE) radiative transfer calculations, based on a previous model (radiation-driven fountain torus) for the Circinus Galaxy (*45*). This possible process complicates our interpretation, so we do not discuss the CO($J = 3 \rightarrow 2$) dynamics in this innermost region. The inclination angle increases gradually toward the center, which is consistent with the values estimated from the $H_2O$ maser observations (*24*) and mid-infrared interferometric observations of dust (*50-52*) observations. The values of $V_{rot}$ we find at $r > 10$ pc are consistent with previous models (Figure S7).

**V. Infall motion observed in the HCN($J = 3 \rightarrow 2$) profile**

We analyzed the inverse P-Cygni profile of the HCN($J = 3 \rightarrow 2$) spectrum, an indication of inflow motion, using a two-layer slab model (*53*) that accounts for the continuum source (*34*). The model assumes that the central continuum source is optically thick and constitutes a fraction of the beam $f$. Both the front and rear layers fall toward this continuum source (the AGN) with an inflow velocity and slab velocity dispersion of $V_{in}$ and $\sigma_s$, respectively. We do not include heating by the cosmic microwave background radiation. The expected line flux density $F(V)$ at velocity $V$ is

$$F(V) = [S_f - fS_c - (1-f)S_r] \times (1-\exp(-\tau_f)) + (1-f)S_r \times (1-\exp(-\tau_f - \tau_r)) \text{ mJy}, \quad (S2)$$

where $S_f$, $S_c$, and $S_r$ denote the source functions of the front layer, continuum source, and rear layer convolved with the beam, respectively. The opacities of the front ($\tau_f$) and the rear ($\tau_r$) layers and their peak values ($\tau_0$) are:

$$\tau_f = \tau_0 \times \exp(-(V-(V_{sys} + V_{in}))^2/2\sigma_s^2), \quad (S3)$$
$$\tau_r = \tau_0 \times \exp(-(V-(V_{sys} - V_{in}))^2/2\sigma_s^2). \quad (S4)$$

For the continuum emission, we used the simultaneously taken continuum map at 1.1 mm wavelength (Figure S8A) to constrain the relevant parameters. We applied the CASA `imfit` task, which indicates that the beam-deconvolved (intrinsic) size of the central brightest region of the continuum emission is $(17.5 \pm 1.5) \times (7.5 \pm 3.2)$ milli-arcsec$^2$, which occupies 20% of the HCN($J = 3 \rightarrow 2$) beam. Thus, we set $f = 0.2$. For simplicity, we set $fS_c$ to the observed peak continuum flux density of 13.5 mJy. To remove potential contamination of foreground emission, we measured an averaged spectrum of two positions located 2 pc away from the center, along the kinematic minor axis of the HCN disk. This averaged spectrum was subtracted from the HCN($J = 3 \rightarrow 2$) raw spectrum, yielding the foreground corrected spectrum. We fitted the model to the HCN($J = 3 \rightarrow 2$) data in a velocity range of 350-540 km s$^{-1}$, to avoid possible contamination of the fast Keplerian rotation seen in emission. We find $S_f = 8.8 \pm 1.1$ mJy, $S_r = 8.6 \pm 1.9$ mJy, $\tau_0 = 2.0 \pm 0.6$, $V_{in} = 7.4 \pm 1.0$ km s$^{-1}$, and $\sigma_s = 41 \pm 3$ km s$^{-1}$. $\sigma_s$ denotes the dispersion measured along the entire length of the foreground (or rear) layer, which is not necessarily identical to the local velocity dispersion of an individual cloud.

With this infall velocity ($V_{in} = 7.4$ km s$^{-1}$) and the estimated disk scale height ($h \sim 0.1 \times r$; Figure 2), we estimate the mass inflow rate as

$$\dot{M}_{in}(r) = 2\pi r \times 2h \times 2n_{H2} \times m_p \times V_{in}, \quad (S5)$$

where $m_p$ denotes the mass of the proton. As the molecular hydrogen volume density is unknown, we used the critical density of HCN($J = 3 \rightarrow 2$), which is $(5.8-10) \times 10^6$ cm$^{-3}$ across the temperature range of 20 to 50 K. A low temperature of ~30 K is typically found in the densest mid-plane gas (shielded region from AGN radiation) of the CNDs in high-resolution hydrodynamic simulations (*33, 54*). At $r = 0.27$ pc (half the geometrical mean of the beam major and minor axes), we obtained an inflow rate of ~0.20–0.34 M$_\odot$ yr$^{-1}$. The black hole accretion rate was estimated from the AGN bolometric luminosity of the Circinus Galaxy (*55*), $L_{Bol} = 4.3 \times 10^{43}$ erg s$^{-1}$. Assuming a radiative efficiency of 10% for the accretion disk (*36*), we obtain an accretion rate of 0.006 M$_\odot$ yr$^{-1}$, which is considerably smaller (< 3%) than the inflow rate estimated above.

Although the inflow velocity was estimated with the absorption profile measured at the AGN position alone, we can roughly estimate the radial profile of the inflow rate by assuming that the inflow velocity is constant across the HCN($J = 3\rightarrow2$) disk. We use $\sigma_{\rm disp} / V_{\rm rot}$ as a proxy for the aspect ratio of the disk. Using Equation (S5), we then obtain the profile shown in Figure S9, for the two gas densities mentioned above. We find an increasing trend of the inflow rate toward the outer radii, though the individual values are quite uncertain because the large uncertainty in $\sigma_{\rm disp}$ propagates to the uncertainty in $\sigma_{\rm disp} / V_{\rm rot}$. Nevertheless, we speculate that these inflow rates, particularly at $r > 1$ pc, are overestimated, for several reasons. Firstly, the actual gas density would be smaller because the HCN($J = 3\rightarrow2$) emission becomes fainter in this outer region, suggesting that the gas density is well below the critical density. Secondly, at the regions more distant from the SMBH, we expect slower inflow velocities (*19*). Thirdly, we do not expect a higher $\sigma_{\rm disp} / V_{\rm rot}$ at $r > 1$ pc than that at the innermost sub-pc region ($\sigma_{\rm disp} / V_{\rm rot} \sim 0.1$). These effects each reduce the inflow rate compared to the values shown in Figure S9.

In the main text, we argued that the absorption due to gas clouds on elliptical orbits is unlikely to explain the HCN($J = 3\rightarrow2$) profile. We base this conclusion on the cloud's mean free path ($L_{\rm mean}$). Consider a gas disk with a radius $r$ is fragmented to $N_{\rm cl}$ clouds by gravitational instability with non-circular motions. The size of these clouds would be within an order of magnitude of the disk scale heigh $h$. Hence $N_{\rm cl} \sim (r/h)^2$. We estimate the mean free path from $(N_{\rm cl}/V) \times L_{\rm mean} \times h^2 \sim 1$, where $V$ is the volume of the disk. By substituting $V = \pi r^2 h$ and the expression for $N_{\rm cl}$, we obtain $L_{\rm mean} \sim 3h$ or $0.3r$ at the central ~1 pc region or inward (Figure 2E). This means that each cloud cannot survive for one orbital period, so the cloud system must be converted to a smooth disk with density fluctuations (*35*).

### VI. Ionized outflow

The derivation of the ionized gas outflow rate requires estimation of the electron density ($n_{\rm e}$), spatial extent of the outflow, and the outflow velocity. The emissivity of the H36α transition over a wide range of $n_{\rm e}$ and electron temperature ($T_{\rm e}$) has been computed under the Case-B recombination condition in previous work (*56*). However, previous studies have questioned whether the Case B condition holds in the vicinity of AGNs (*57, 58*). The departure coefficient of H36α from the LTE condition must approach unity (*56*) for a wide range of $n_{\rm e}$ and $T_{\rm e}$, considering the small energy gap between the principal quantum numbers of $n = 36$ and 35. We therefore estimated $n_{\rm e}$ assuming LTE conditions. The photoionization of the coronal lines (*27*) requires $T_{\rm e}$ of ~$10^4$ K. Previous observations of [S II] and [S III] lines indicate ($n_{\rm e}$, $T_{\rm e}$) = (516 cm$^{-3}$, 10500 K) at the central ~16 pc of the ionization cone of the Circinus Galaxy (*26*). Comparing this temperature (10500 K) with the measured line peak intensity 1.3 mJy (20 K), we determine a line optical depth of $\tau_{\rm H36\alpha} = 1.9 \times 10^{-3}$. This opacity can be analytically described as

$$\tau_{\rm LTE} = 2.7 \times 10^3 \times (T_{\rm e}/{\rm K})^{-3} \times (EM/{\rm cm^{-6}\ pc}) \times (\nu/{\rm GHz})^{-1}, \quad {\rm (S6)}$$

where $EM$ denotes the emission measure along the path and ν is the frequency. We thus obtained $EM = 1.1 \times 10^8$ cm$^{-6}$ pc. We modeled the H36α distribution as a conical horn with a 1 pc base-radius and 2 pc height (Figure 1D) and assumed a line-of-sight path length of 2 pc. This provides an estimate of $n_{\rm e}$ ~7400 cm$^{-3}$.

Therefore, the total ionized gas mass inside this conical volume ($2\pi/3$ pc$^3$) is 383 M$_\odot$. For the outflow velocity ($V_{\rm out}$), we adopted the prescription (*59, 60*) of 0.5 × line FWHM = 197 km s$^{-1}$ (Table S2). The flow travel time from the center to the 2 pc length at this velocity is 9934 yr, resulting in the expected ionized outflow rate of ~0.04 $M_\odot$ yr$^{-1}$.

This ionized outflow rate is consistent within uncertainties with the values measured at ~200 pc from the center by using the optical Hβ line (*26*), indicating that this nuclear wind extends to a larger scale. We computed the outflow kinetic power and the momentum load (normalized by the radiative momentum of the AGN) as

$$\dot{E}_{out} = \frac{1}{2} \dot{M}_{out} v_{out}^2, \qquad (S7)$$

$$\dot{P}_{out}/\dot{P}_{AGN} = \frac{\dot{M}_{out} v_{out}}{L_{Bol}/c}, \qquad (S8)$$

respectively. We obtain $\dot{E}_{out}$ as ~5 × 10$^{38}$ erg s$^{-1}$, which corresponds to ~0.001% of the AGN bolometric luminosity, and $\dot{P}_{out}/\dot{P}_{AGN}$ of 0.035. Both of these values are considerably smaller than predicted by models of AGN feedback onto the host galaxy, e.g., ~5% and 20% in the energy-conserving mode of AGN feedback (*20*). Thus, it seems unlikely that this ionized wind has an impact on the host galaxy star formation.

There is a spatial relation between H36α and dust continuum emission (Figure S8B). As observed, the 1.1-mm-continuum distribution is complex, with a bright core and an extended component (length ~ 2 pc on both the northwest and southeast sides, PA ~ 295°). Although the extended component is almost symmetric in relation to the AGN position, it is not smooth. The spatial distribution of this extended 1.1 mm continuum emission is consistent with the previously observed mid-infrared (MIR) dust continuum distributions (*51*), implying that this 1.1 mm emission traces the submillimeter counterpart of the dust in the disk polar direction (*50, 52*).

The bright region of this extended 1.1-mm-continuum emission (as well as that of the MIR continuum emission) corresponds to ridge-2 of the H36α distribution (Figure S8B). Because H36α traces the ionized outflow, we suggest that the extended dust emission (i.e., polar dust) originates from the rim of the outflowing dusty cone. This direction is consistent within uncertainties with that of the X-ray Fe-Kα extension (*47, 61*). Channel maps show bright [C I] $^3P_1$–$^3P_0$ also in this direction (Figure S3). The X-ray radiation, escaping toward the direction of least obscuration, could also cause a X-ray dissociation of molecular gas in this direction. The continuum is not bright at ridge-1, but this could be due to asymmetric illumination of the dusty cone by the AGN (*52*), or inhomogeneous density structures there. We do not detect H36α emission from the southeast side of the AGN. We suggest this is due to insufficient sensitivity. As submm recombination lines are typically optically thin, their luminosities are proportional to the EM. Hence if there is a factor of 2 variation in the local gas density between the northwest and the southeast sides, the emission from the latter could fall below the 3σ detection level of our data.

**VII. [C I] $^3P_1$–$^3P_0$ outflow and its contribution to the AGN obscuration**
The two [C I] $^3P_1$–$^3P_0$ peaks in the kinematic minor axis PVD are offset by ~15 km s$^{-1}$ from $V_{sys}$. The southeast side of the Circinus Galaxy is the near side (*11*), so the velocity offset distribution (components at $V_{LOS} > V_{sys}$ mostly appear at the positive spatial offset, whereas $V_{LOS} < V_{sys}$ appears in the opposite direction) indicates the presence of outflows. The fast ($V_{LOS}$ ~300 km s$^{-1}$) component along this minor axis PVD corresponds to the Keplerian rotation caused by the central SMBH (*24*) convolved with the synthesized beam.

Using the inclination angle of the CND (*i*), the maximum velocity offset corresponds to
$$|V_{LOS} - V_{sys}| = V_{out} \times \cos(\varDelta + \pi/2 - i), \qquad (S9)$$

where $V_{out}$ denotes the outflow velocity, and $\Delta$ indicates the launching angle of the conical outflow measured from the disk plane. The value of $\Delta$ is unknown, so we estimate it from the half-opening angle ($\varphi$) of the ionization cone traced by H36α. Assuming that atomic carbon is not present within this ionization cone, we obtain $\Delta < \pi/2 - \varphi$. Based on the kinematically estimated $i$ of ~80° (Figure S7) and $\varphi$ measured from our H36α observation (33°–39°, Figure 1D), we obtain $V_{out} < 38$ km s$^{-1}$. This outflow velocity is smaller than the escape velocity from the central SMBH potential, at $r = 1$ to 3 pc region (which is ~70 to 120 km s$^{-1}$ for our measured black hole mass). The outflow will thus become a backflow to the disk.

The [C I] $^3P_1$–$^3P_0$ line flux measured along the kinematic minor axis is 5.0 to 9.1 Jy beam$^{-1}$ km s$^{-1}$ at r < 5 pc surrounding the nucleus (Figure 1A), corresponding to a beam-enclosed line luminosity of $L'_{[CI]} = (1.2–2.2) \times 10^4$ K km s$^{-1}$ pc$^2$. We use this to estimate the atomic carbon mass ($M_{CI}$) following previous methods (62, 63) as

$$M_{CI} = 5.71 \times 10^{-4}\ Q(T_{ex})\ (1/3)\ \exp(23.6/T_{ex})\ L'_{[CI]}\ M_\odot, \qquad (S10)$$

where $Q(T_{ex}) = 1 + 3\exp(-T_1/T_{ex}) + 5\exp(-T_2/T_{ex})$ denotes the C I partition function, $T_1 = 23.6$ K and $T_2 = 62.5$ K represent the energy levels above the ground state. This analysis assumes LTE and optically thin conditions. The line excitation temperature is unknown, so we consider the range $T_{ex} = 100–1000$ K (line peak intensity of ~120 mJy beam$^{-1}$ corresponds to the brightness temperature of ~70 K, and thus, the excitation temperature should be greater than this value) which yields $M_{CI} = 17–36\ M_\odot$, as measured with the single synthesized beam. After assuming the typical atomic carbon-to-atomic hydrogen abundance ratio of $X_{CI} = 10^{-4}$ found by XDR chemical simulations (30, 64), we calculated a hydrogen mass of $M_H = M_{CI}/X_{CI} \times (1/12) = (1.4–3.0) \times 10^4\ M_\odot$, corresponding to the line-of-sight hydrogen column density of $N_H = (4.2–8.8) \times 10^{23}$ cm$^{-2}$.

Assuming the typical relation between the extinction in the optical $V$-band and $N_H$ in the Milky Way (65) $N_H = 2 \times 10^{21} \times A_V$ cm$^{-2}$ also applies to the Circinus Galaxy, we obtained a visual extinction of $A_V = 210–440$ mag. We assumed that the dust particles survive in this atomic outflow, as found by a previous chemo-radiation hydrodynamic simulation of the Circinus galaxy, which modeled the gas and dust structures at the central ~10 pc scale (33). In most regions at $r > 0.1$ pc, the dust temperature does not exceed the sublimation temperature (~1500 K). Sputtering by the hot gas is not effective as the gas density is very low in that phase. That simulation also showed that the atomic carbon and dust can be co-spatial near the outer edge of the ionization cone, i.e., near the inner wall of the dust torus. As the inclination angle of the central pc-scale region of the Circinus galaxy is close to 90°, the line flux measured along the minor axis implies that we measured the flux and extinction of the components located above the central disk, as observed from the nearly edge-on view. Therefore, we conclude that this structure causes substantial obscuration, as expected for an AGN torus (33, 66).

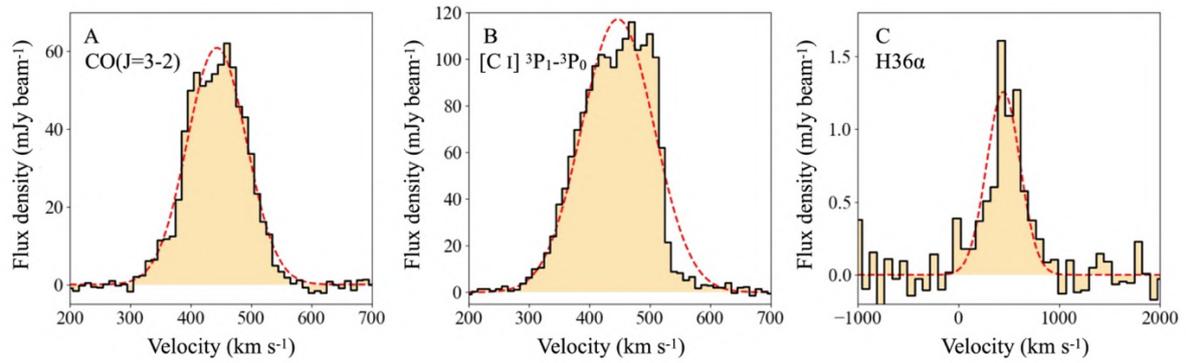

**Figure S1. Line spectrum measured at AGN position.** Same as Fig. 3A, but for the **(A)** CO($J = 3\rightarrow2$), **(B)** [C I] $^3P_1$–$^3P_0$, and **(C)** H36α emission. Dashed red lines indicate single Gaussian models fitted to the data (parameters listed in Table S2).

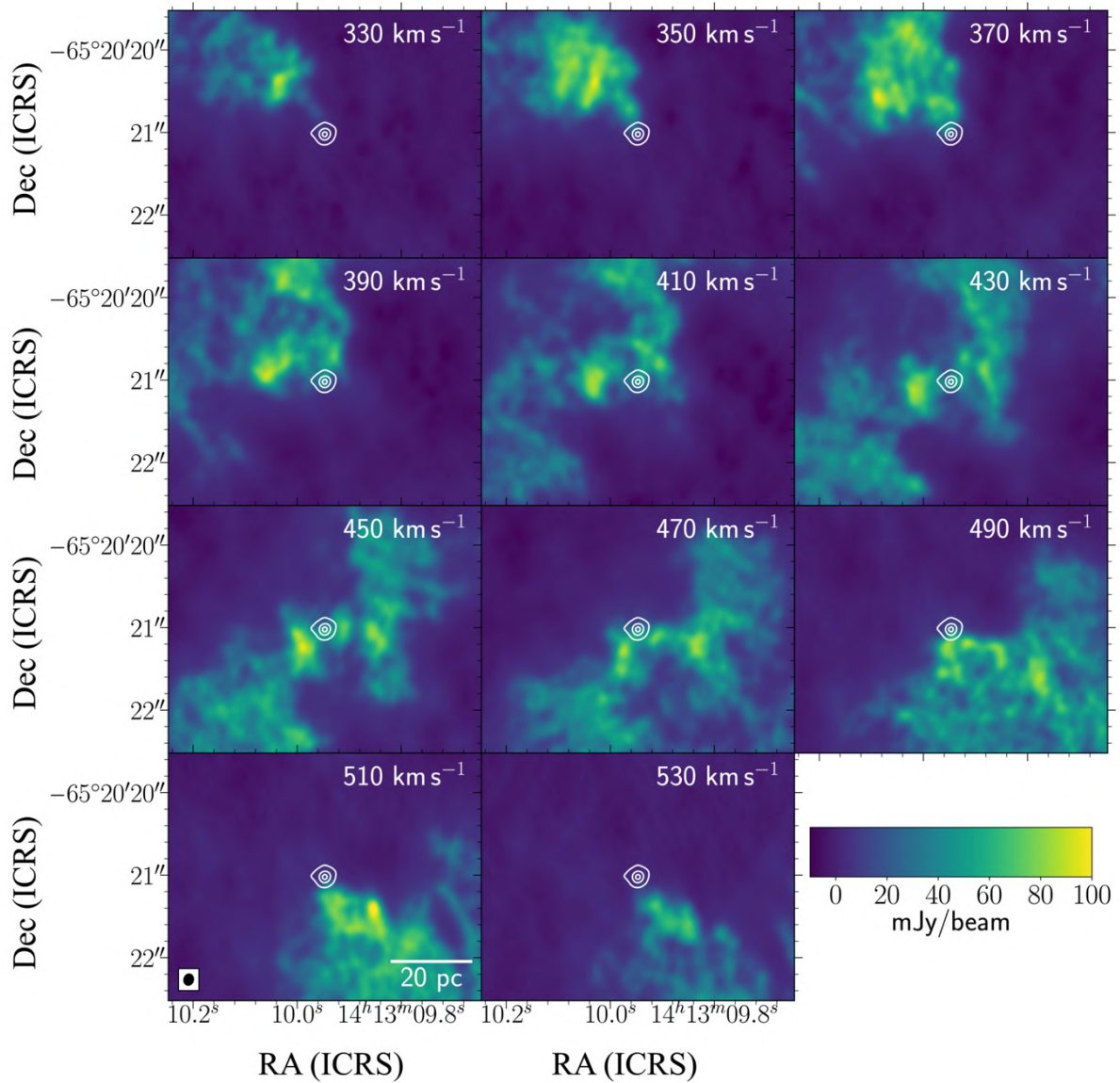

**Figure S2. Velocity channel maps of CO(*J* = 3→2).** The central 3.0 arcsec region of the Circinus galaxy is shown. The contours indicate the underlying 860 μm continuum distribution, and are drawn at 20, 70, and 140σ (1σ = 71 μJy beam$^{-1}$). Different panels indicate the flux density distributions in different velocity channels, indicated at the top-right corner of each panel. In each panel, colors indicate the observed flux density (mJy beam$^{-1}$). The black ellipse on a white square indicates the synthesized beam.

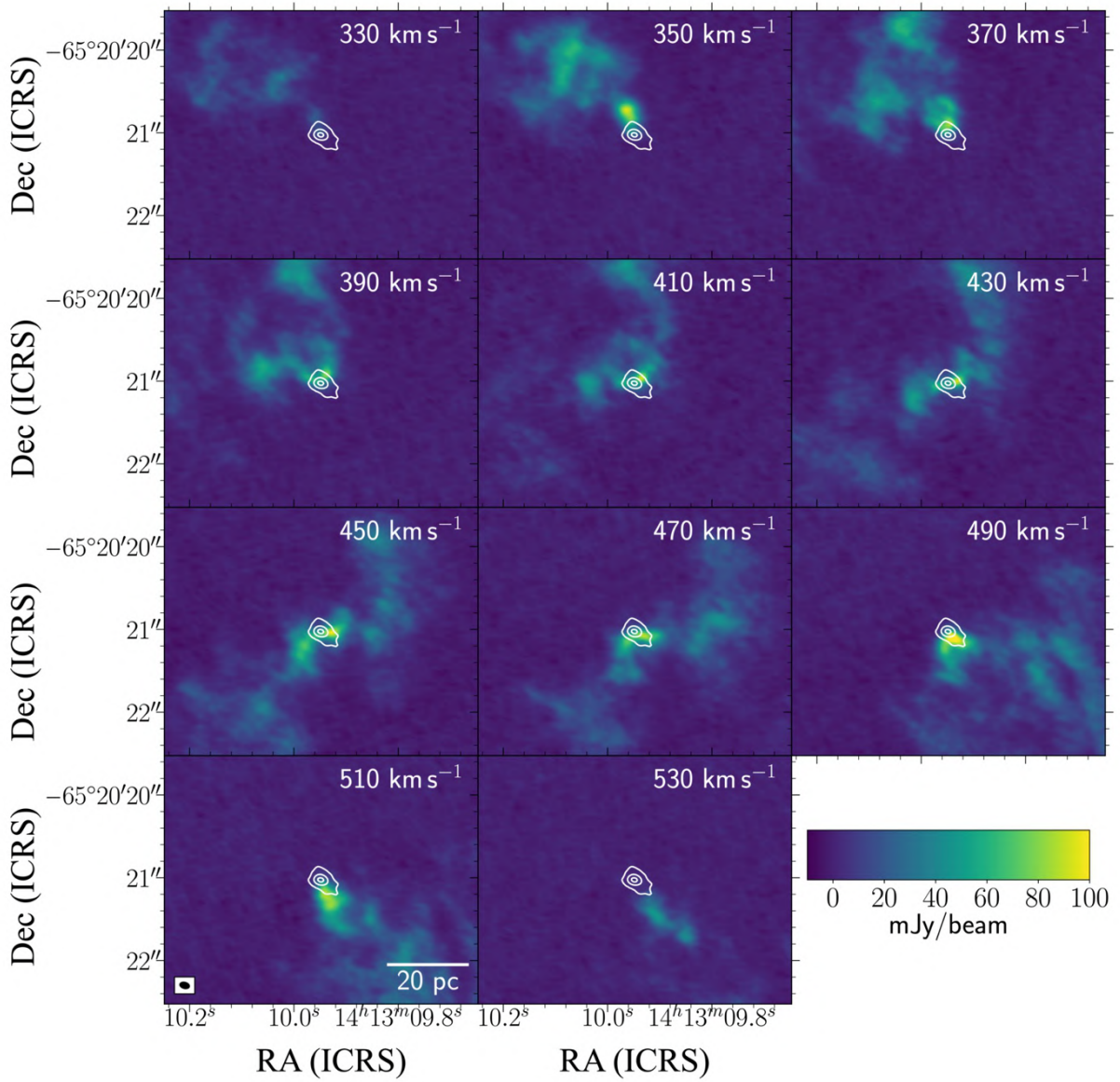

**Figure S3. Velocity channel maps of [C I] $^3P_1$–$^3P_0$.** Same as Fig. S2 but for the atomic carbon observations. The contours indicate the underlying 600 μm continuum distribution, which are drawn at 20, 70, and 140σ (1σ = 91 μJy beam$^{-1}$).

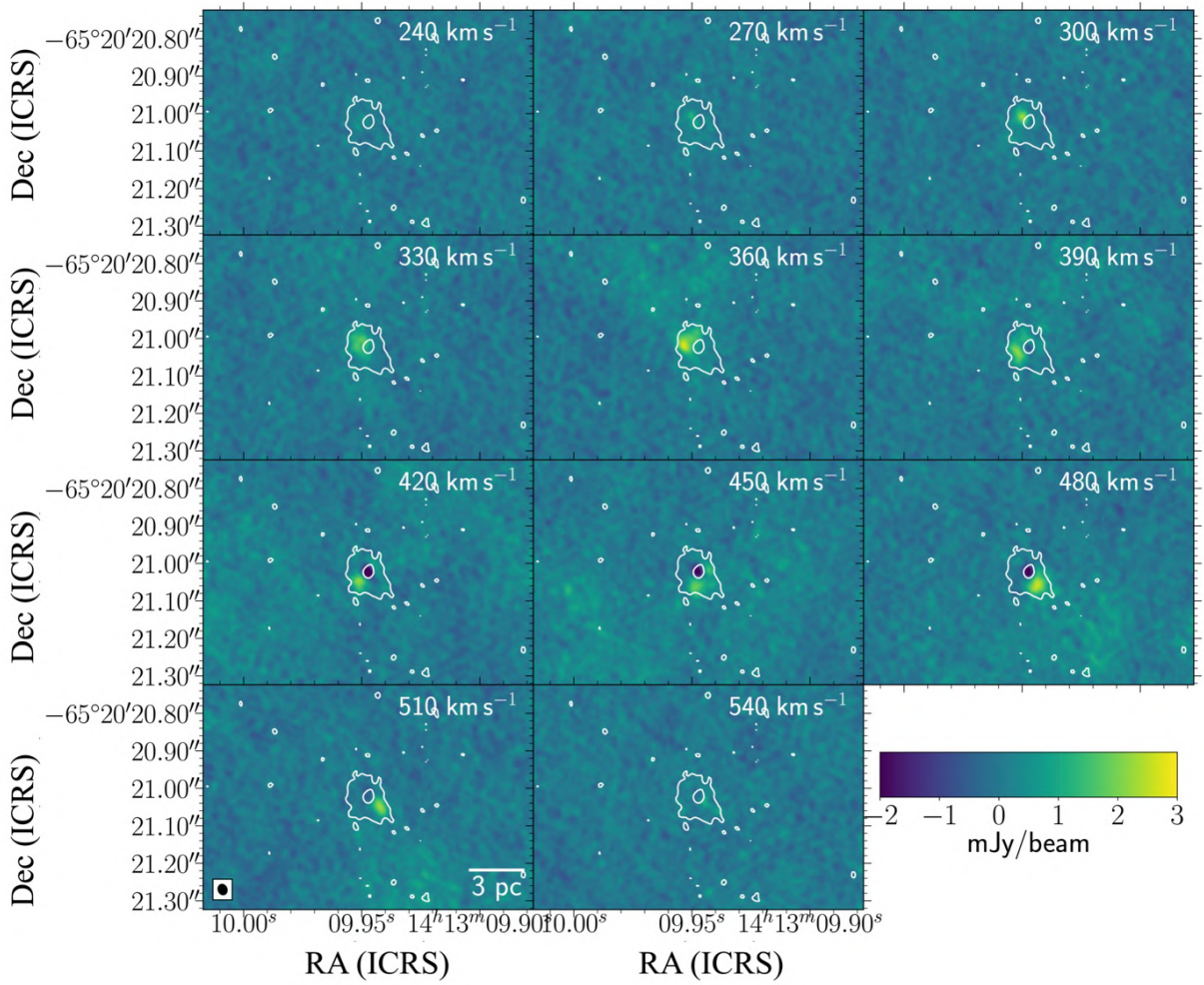

**Figure S4. Velocity channel maps of HCN($J = 3 \rightarrow 2$).** Same as Fig. S2, but for the HCN observations. A larger 0.6 arcsec region is shown. The contours indicate the 5σ distribution of total HCN($J = 3 \rightarrow 2$) integrated intensity over all velocities.

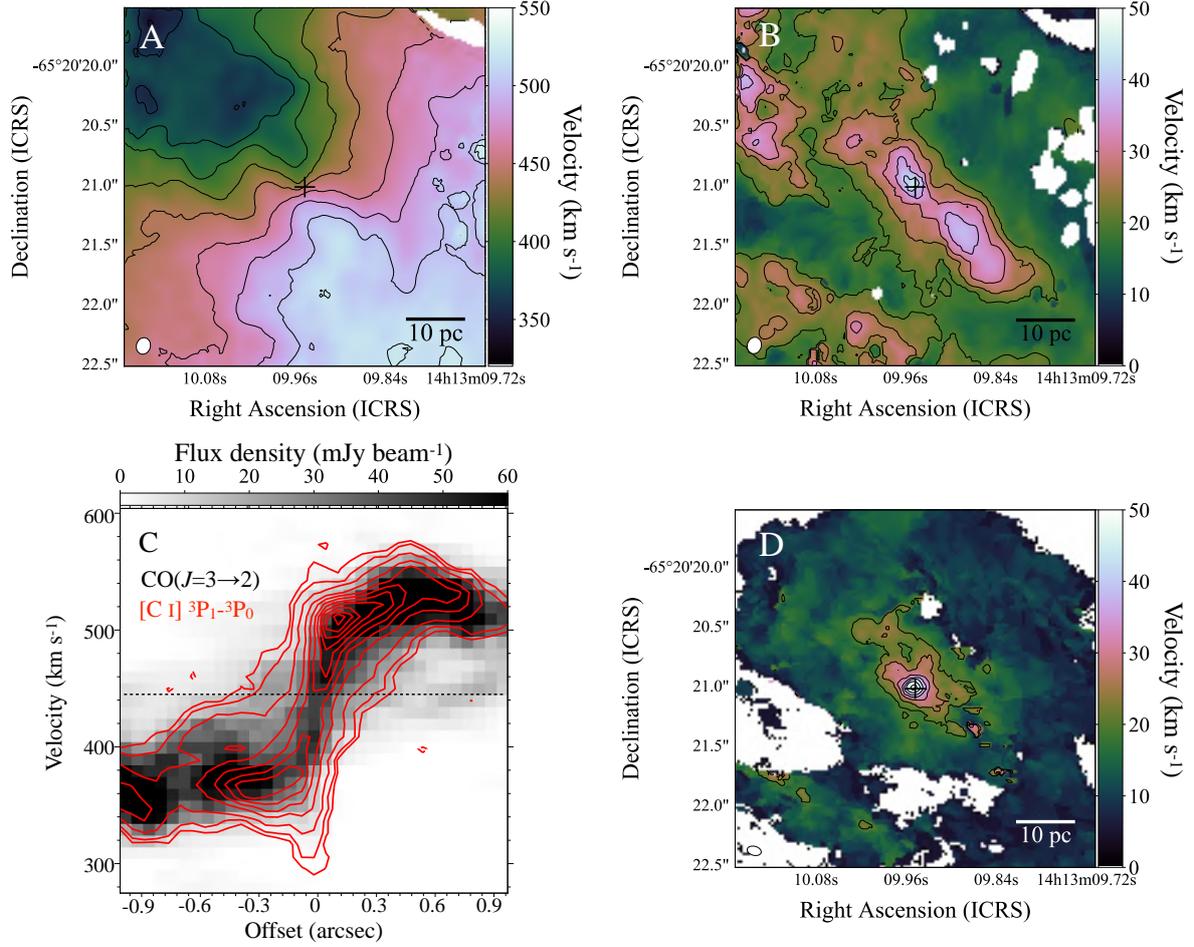

**Figure S5. Velocity maps of CO and [C I] on larger scales.** (**A**) CO($J = 3\rightarrow2$) line-of-sight velocity map (color-scale). Contours are drawn from 360 km s$^{-1}$ in steps of 20 km s$^{-1}$ ($V_{\rm sys}$ = 446 km s$^{-1}$). (**B**) CO($J = 3\rightarrow2$) velocity dispersion map (color-scale). Contours increase from 20 km s$^{-1}$ in steps of 5 km s$^{-1}$. (**C**) Major axis PVDs of CO($J = 3\rightarrow2$) (grayscale indicate the flux density, 1$\sigma$ = 0.18 mJy beam$^{-1}$) and [C I] $^3P_1$–$^3P_0$ (red contours drawn at 3$\sigma$, 5$\sigma$, 10$\sigma$, 20$\sigma$, …, 90$\sigma$, where 1$\sigma$ = 1.26 mJy beam$^{-1}$). (**D**) [C I] $^3P_1$–$^3P_0$ velocity dispersion map (color-scale). Contours increase from 20 km s$^{-1}$ in steps of 5 km s$^{-1}$. Although the line-of-sight velocity map of [C I] $^3P_1$–$^3P_0$ is consistent with that of CO($J = 3\rightarrow2$) (Fig. 2), the dispersion patterns are different.

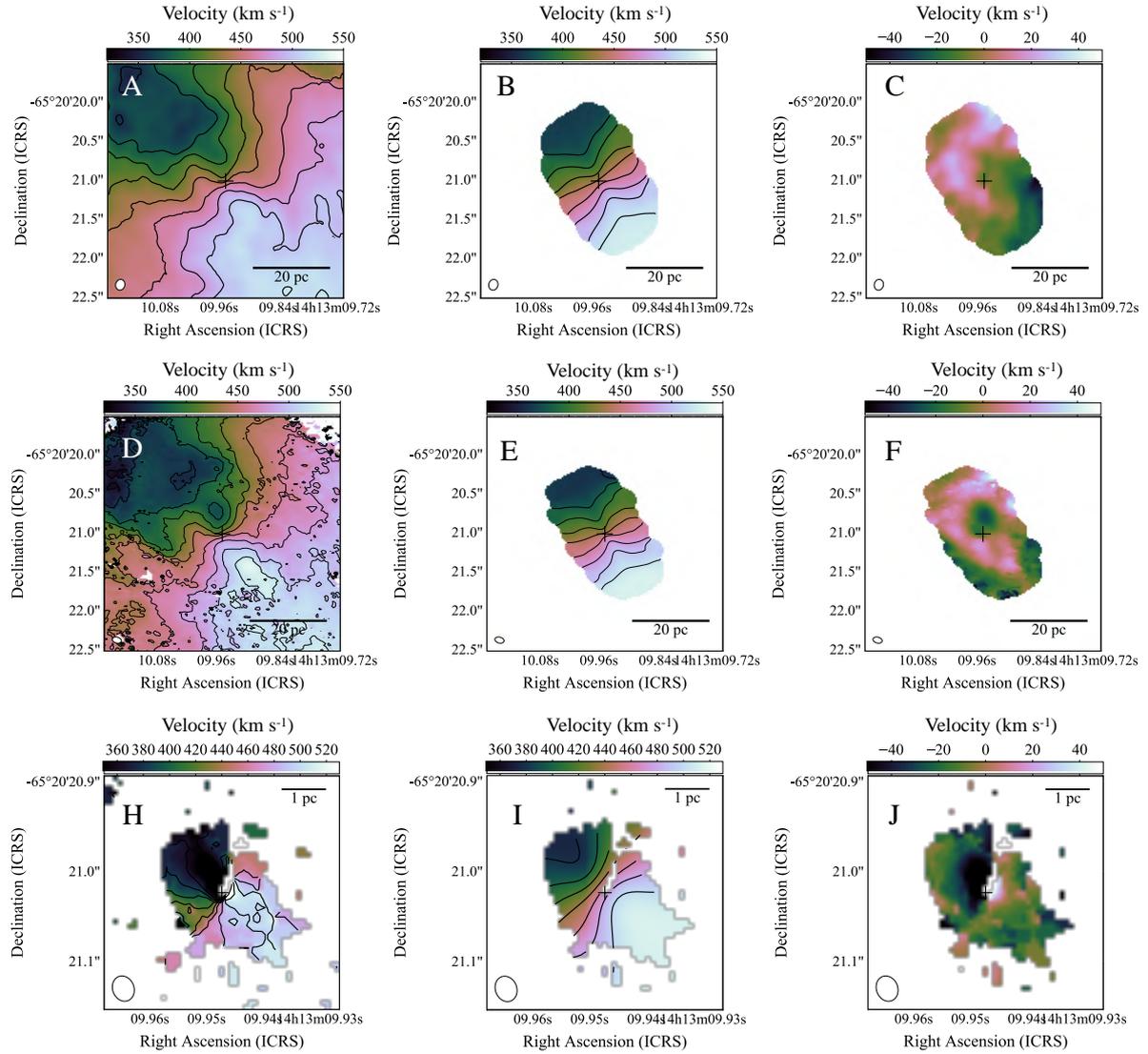

**Figure S6. Dynamical models constructed for the Circinus galaxy.** (**A**) Observed line-of-sight velocity field, (**B**) modeled velocity field, and (**C**) residual computed by subtracting the model from the observation, all for CO($J = 3{\rightarrow}2$). Colors indicate the velocities for these components. In panels A and B, contours start from 320 km s$^{-1}$ and increase in steps of 20 km s$^{-1}$. (**D**)-(**F**) Equivalent maps for [C I] $^3P_1$–$^3P_0$. In panels D and E, contours start from 320 km s$^{-1}$ and increase in steps of 20 km s$^{-1}$. (**G**)-(**I**) Equivalent maps for HCN($J = 3{\rightarrow}2$). In panels H and I, contours start from 360 km s$^{-1}$ and increase in steps of 20 km s$^{-1}$. In each panel, the central plus sign indicates the AGN position. The black ellipses in the bottom left of each panel indicate the synthesized beams. In all cases, the residual is close to 0 km s$^{-1}$ close to the AGN.

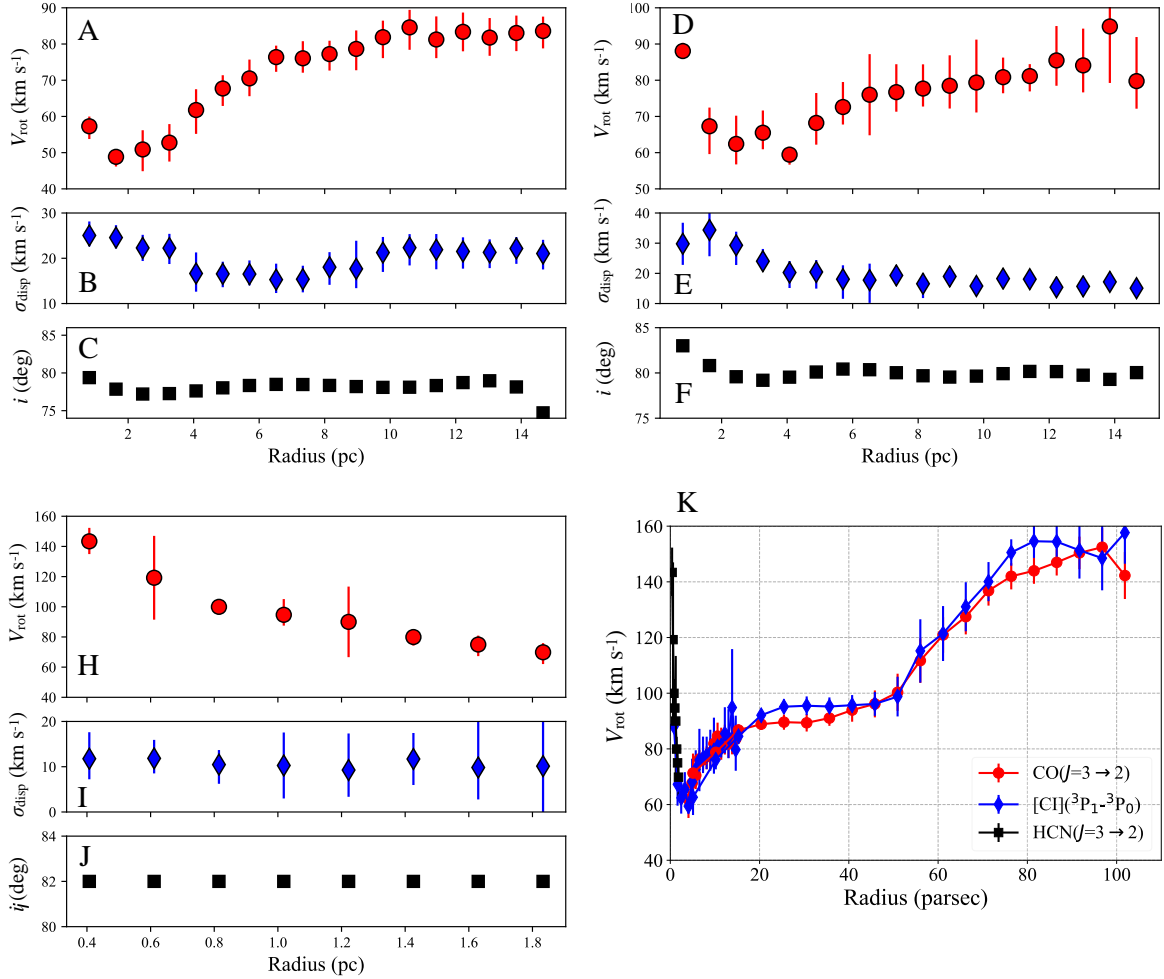

**Figure S7. Dynamical parameters from the models.** Rotation velocity ($V_{rot}$, red circles), velocity dispersion ($\sigma_{disp}$, blue diamonds), and inclination angle ($i$, black squares) for the cases **(A)** CO($J = 3\rightarrow2$), **(B)** [C I] $^3P_1$–$^3P_0$, and **(C)** HCN($J = 3\rightarrow2$). **(D)** Combination of our model rotation curves and a previous large-scale rotation curve constructed by using CO($J = 3\rightarrow2$) and [C I] $^3P_1$–$^3P_0$ at $r > 20$ pc (*11*), shown from 100 pc to the central subpc-scale. Error bars indicate 1σ uncertainties.

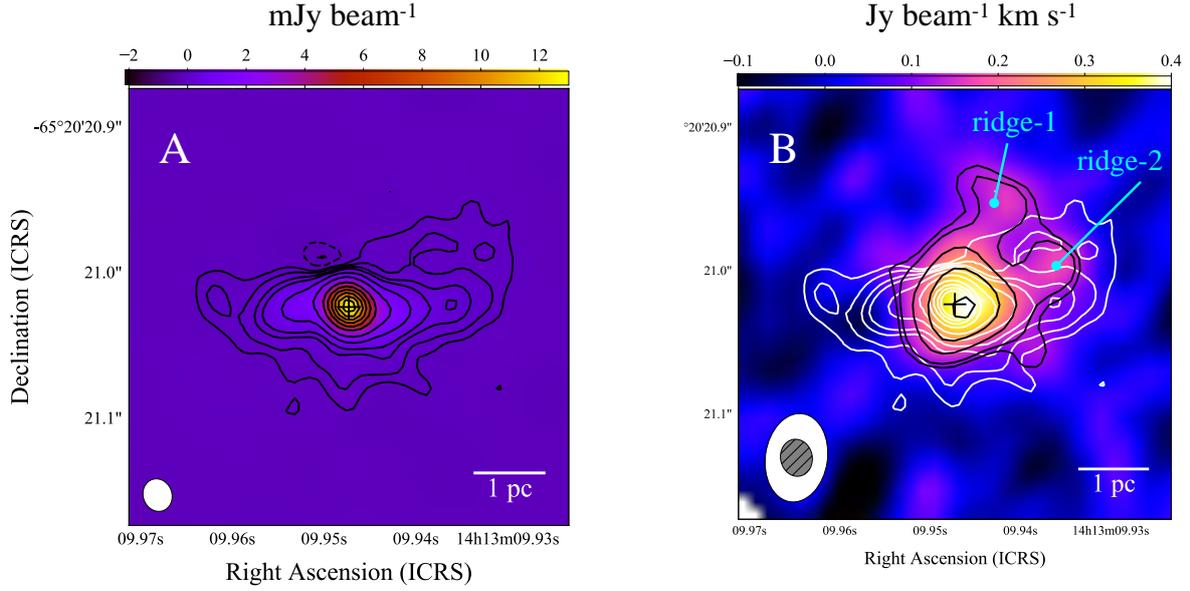

**Figure S8. Band 6 continuum emission distribution. (A)** 1.1 mm continuum emission map (color scale) of the central region of the Circinus Galaxy. Contours are drawn at -10, -5, 5, 10, 20, 30, 60, 120, 240, 360, 480, 600, 720, 840, and 960σ, where 1σ = 14.7 μJy beam$^{-1}$. The beam-deconvolved size of the central bright portion is (17.5 ± 1.5) × (7.5 ± 3.2) milli-arcsec$^2$ with PA = 81.4°. **(B)** The H36α distribution (color-scale and black contours, contours are drawn at 3, 3.5, 5, 7, and 10σ, where 1σ = 0.040 Jy beam$^{-1}$ km s$^{-1}$) compared to the 1.1 mm continuum emission (white contours, drawn at the same levels as panel A). The ellipses in the bottom left indicate the synthesized beam of the H36α data cube (white) and the 1.1 mm continuum map (gray with hatching), respectively.

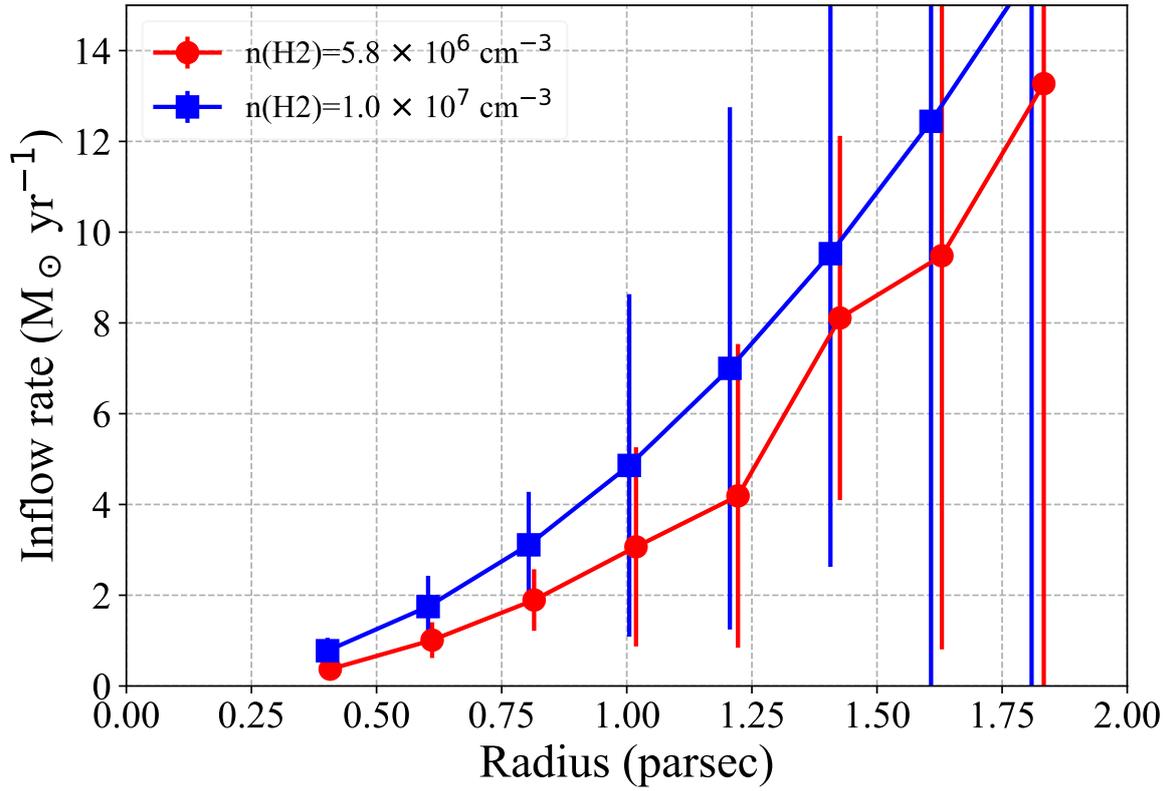

**Figure S9 Radial profile of mass inflow rate estimated from the HCN($J = 3\rightarrow2$) dynamics.** We assumed two gas densities (red circles = $5.8 \times 10^6$ cm$^{-3}$, blue squares = $1.0 \times 10^7$ cm$^{-3}$) and that the inflow velocity is constant (7.4 km s$^{-1}$) across the HCN disk. The two profiles have been slightly shifted for display (blue squares are placed 1.5% to the left of the red circles). Error bars indicate 1σ uncertainties. There is an increasing trend toward larger radii. However, we expect the actual inflow rate at larger radii to be much smaller, due to the uncertainty in gas dynamics and local gas volume density (see text).

**Table S1. Data cube parameters for the observations of different target emission lines.**

| Line Name | Rest frequency (GHz) | Angular resolution (arcsec × arcsec) | Velocity resolution (km s$^{-1}$) | 1σ sensitivity (mJy beam$^{-1}$) |
|---|---|---|---|---|
| HCN($J = 3{\rightarrow}2$) | 265.8864 | 0.029 × 0.024 | 15 | 0.20 |
| H36α high | 135.2984 | 0.062 × 0.043 | 75 | 0.18 |
| H36α low | 135.2984 | 0.058 × 0.041 | 75 | 0.18 |
| CO($J = 3{\rightarrow}2$) | 345.7960 | 0.140 × 0.117 | 10 | 0.78 |
| [C I] $^3P_1$–$^3P_0$ | 492.1607 | 0.119 × 0.076 | 10 | 1.26 |

**Table S2. Results from fitting Gaussian models to the line spectra.** HCN($J = 3{\rightarrow}2$) is not included because it has an absorption feature (Fig. 3A). The line flux was calculated by integrating the Gaussian profile. We also list line fluxes measured with moment-0 maps for comparison.

| Line | Center (km s$^{-1}$) | Peak (mJy beam$^{-1}$) | FWHM (km s$^{-1}$) | Flux (Jy beam$^{-1}$ km s$^{-1}$) | Flux, mom-0 (Jy beam$^{-1}$ km s$^{-1}$) |
|---|---|---|---|---|---|
| CO($J = 3{\rightarrow}2$) | 443.3 ± 0.3 | 60.9 ± 0.3 | 115.2 ± 0.8 | 7.5 ± 0.6 | 7.3 ± 0.1 |
| [C I] $^3P_1$–$^3P_0$ | 447.1 ± 0.4 | 117.1 ± 0.6 | 141.1 ± 0.8 | 17.6 ± 1.0 | 17.4 ± 0.1 |
| H36α | 446.0* | 1.3 ± 0.1 | 393.1 ± 44.2 | 0.5 ± 0.1 | 0.4 ± 0.1 |

*Fixed to the systemic velocity.